\documentclass[a4paper,11pt]{article}
\usepackage{graphicx,amssymb,amstext,amsmath,mathabx}
\usepackage{color, yfonts, tcolorbox}
\usepackage{floatflt}

\usepackage{float}
\usepackage{esint}
\usepackage{subcaption}
\usepackage{cite}
\usepackage{mathtools}
\usepackage{commath}
\usepackage{array}
\usepackage{boldline}
\usepackage{multirow}
\usepackage{arydshln}
\usepackage{bigdelim}
\usepackage{hyperref}
\definecolor{cbl}{rgb}{0,0,1}

\DeclareMathOperator{\sech}{sech}

\newcommand{\bc}{\begin{center}}
\newcommand{\ec}{\end{center}}
\def\ba#1{\begin{array}{#1}\displaystyle}
\newcommand{\ea}{\end{array}}

\newcommand{\beq}{\begin{equation}}
\newcommand{\eeq}{\end{equation}}
\newcommand{\beqa}{\begin{eqnarray}}
\newcommand{\eeqa}{\end{eqnarray}}

\newcommand{\bi}{\begin{itemize}}
\newcommand{\ei}{\end{itemize}}

\newcommand{\bra}{\langle}
\newcommand{\ket}{\rangle}

\newcommand{\TT}{{\cal T}}

\begin{document}
\begin{titlepage}
\vspace{0.2cm}
\begin{center}

{\large{\bf{Two-Point Functions of Composite Twist Fields in the Ising Field Theory}}}

\vspace{0.8cm} 
{\large   Olalla A. Castro-Alvaredo and Michele Mazzoni}

\vspace{0.8cm}
{\small
 Department of Mathematics, City, University of London, 10 Northampton Square EC1V 0HB, UK\\
\medskip
}
\end{center}

\medskip
\medskip
\medskip
\medskip

All standard measures of bipartite entanglement in one-dimensional quantum field theories can be expressed in terms of correlators of branch point twist fields, here denoted by $\TT$ and $\TT^\dagger$. These are symmetry fields associated to cyclic permutation symmetry in a replica theory and having the smallest conformal dimension at the critical point. Recently, other twist fields (composite twist fields), typically of higher dimension, have been shown to play a role in the study of a new measure of entanglement known as the symmetry resolved entanglement entropy. In this paper we give an exact expression for the two-point function of a composite twist field that arises in the Ising field theory. In doing so we extend the techniques originally developed for the standard branch point twist field in free theories as well as an existing computation due to Horváth and Calabrese of the same two-point function which focused on the leading large-distance contribution. We study the ground state two-point function of the composite twist field $\TT_\mu$ and its conjugate $\TT_\mu^\dagger$. At criticality, this field can be defined  as the leading field in the operator product expansion of $\TT$ and the disorder field $\mu$.  We find a general formula for $\log \bra  \TT_\mu(0) \TT^\dagger_\mu(r)\ket$  and for (the derivative of) its analytic continuation to positive real replica numbers greater than 1. We check our formula for consistency by showing that at short distances it exactly reproduces the expected conformal dimension  

\medskip
\medskip

\vfill

\noindent 
o.castro-alvaredo@city.ac.uk\\
michele.mazzoni.2@city.ac.uk\\

\hfill \today

\end{titlepage}
\section{Introduction}
It is well-known that the Ising field theory has an internal $\mathbb{Z}_2$ symmetry, associated to which we can define two fields: $\sigma$, the spin field (order operator), and $\mu$ (disorder operator)\footnote{In this paper we use the conventions of \cite{YZam}, which corresponds to choosing the disordered phase of the model, where the fields $\sigma (\mu)$ are odd (even) with respect to the Majorana fermion $\Psi$.}. The theory also contains a free Majorana fermion field $\Psi$ so that, in the disordered phase of the theory, the three fields can be characterised by their mutual equal-time exchange relations \cite{ZI,ST,BB}:
\beq
\Psi (x)\sigma(y)=\left\{ \begin{array}{cc}
\sigma(y) \Psi(x)& y>x\\
\sigma (y) \Psi(x) & y<x
\end{array}\right. \quad \mathrm{and}\quad \Psi (x)\mu(y)=\left\{ \begin{array}{cc}
-\mu(y) \Psi(x)& y>x\\
\mu (y) \Psi(x) & y<x
\end{array}\right.
\label{simu}
\eeq 
Furthermore, in the context of the investigation of entanglement measures it is often convenient to consider a ``replica" version of the theory, namely a model consisting of $n$ non-interacting, identical copies of the original. In this model, the fields above acquire an index $\{\mu_j, \sigma_j, \Psi_j\}$ with $j=1,\ldots,n$, running over the copy numbers. 
The resulting model possesses a large amount of symmetry, namely, not only $\mathbb{Z}_2$ symmetry on each copy, but symmetry under the exchange of any copies. Cyclic permutation symmetry is one of these many symmetries and, as it turns out, it is the symmetry that plays the most fundamental role in computations of the entanglement entropy and other measures of entanglement \cite{HolzheyLW94,Calabrese:2004eu,entropy}. 

In \cite{entropy} the branch point twist fields $\TT$ and its conjugate $\TT^\dagger$ (called $\tilde{\TT}$ in the original paper) were defined as the symmetry fields associated with cyclic permutation symmetry of copies in a replica theory. These fields too are characterised by their exchange relations with respect to the fermions:
\beq
\Psi_j (x)\TT(y)=\left\{ \begin{array}{cc}
\TT(y) \Psi_{j+1}(x)& y>x\\
\TT (y) \Psi_j(x) & y<x
\end{array}\right. \quad \mathrm{and}\quad \Psi_j (x)\TT^\dagger(y)=\left\{ \begin{array}{cc}
\TT^\dagger(y) \Psi_{j-1}(x)& y>x\\
\TT^\dagger(y) \Psi_j(x) & y<x
\end{array}\right.
\label{T}
\eeq 
for $j=1, \ldots, n$ and $j\equiv j+n$. These relations can be written for any 1+1D quantum field theory, integrable or not, however, in the context of massive integrable quantum field theory (IQFT), they provide, together with the two-body scattering matrix, all the information needed to compute correlation functions and matrix elements of $\TT$ \cite{entropy}. These computations have now been carried out for many theories and entanglement measures (see e.g. \cite{back, nexttonext, E8,FBoson}) revealing many new insights into the universal properties of entanglement at near-critical points. 

In recent years, it has been shown that also the fields resulting from the conformal OPE of $\TT$ with other fields of the Ising field theory can be of interest in the context of entanglement \cite{ctheorem,Levi, BCDLR, BCD,GS, german3}. In particular, the correlation functions of the leading field in the OPE of $\TT$ and $\sum_j \mu_j$, denoted by $\TT_\mu$, are related to an entanglement measure known as the symmetry resolved entanglement entropy \cite{GS,german3,SymResFF}. $\TT_\mu$ satisfies exchange relations which combine those for $\TT$ and  $\mu$ as seen above:
\beq
\Psi_j (x)\TT_\mu(y)=\left\{ \begin{array}{cc}
-\TT_\mu(y) \Psi_{j+1}(x)& y>x\\
\TT_\mu (y) \Psi_j(x) & y<x
\end{array}\right. \quad \mathrm{and}\quad \Psi_j (x)\TT_\mu^\dagger(y)=\left\{ \begin{array}{cc}
-\TT_\mu^\dagger(y) \Psi_{j-1}(x)& y>x\\
\TT_\mu^\dagger(y) \Psi_j(x) & y<x
\end{array}\right.
\label{Tmu}
\eeq 
In general, such measures can always be defined for theories that possess an internal symmetry (such as $\mathbb{Z}_2$ in the Ising case), and it gives access to information about the amount of entanglement that is stored in each symmetry sector.
\medskip

The computation of the symmetry resolved entanglement provides strong motivation to study correlators of $\TT_\mu$ and this will be the focus of this paper. Our aim is  finding an exact analytic expression for the two-point function $\bra \TT_\mu(0)\TT^\dagger_\mu(r)\ket$ using IQFT techniques. The applications of such a result in the context of entanglement will not be discussed here, but they follow quite straightforwardly from existing literature. In particular, the form factors of $\TT_\mu$ and the leading contribution to its two-point function were computed in \cite{SymResFF}. The present work  is an extension of those results to include higher particle contributions and to show how non-trivial resummation identities allow for relatively simple closed formulae for all correlation function cumulants. 

Correlation functions of composite twist fields have been studied in a number of works  both for the Ising field theory and other, interacting models. Most of these results build upon the form factor program for the matrix elements of $\TT$ \cite{entropy} and its extension to composite twist fields \cite{SymResFF}. In \cite{Bonsignori_2019,Murciano_2020,Horvath_2021} free theories were studied, whereas interacting IQFTs such as the Ising and sinh-Gordon models (both with discrete $\mathbb{Z}_2$ symmetry), the sine-Gordon model (with continuous $U(1)$ symmetry) and the $3$-state Potts model (with discrete $\mathbb{Z}_3$ symmetry) were studied in \cite{SymResFF,horvath2021branch} and  \cite{pottsSRE}, respectively.

It is also possible to study composite twist fields where $\TT$ is composed with a local field {\it not} associated with an internal symmetry. Such composite fields are associated with cyclic permutation symmetry too and have a conformal dimension which is distinct from that of $\TT$. In particular, for theories whose UV fixed point is described by a non-unitary conformal field theory (CFT), it is possible to construct composite twist fields whose dimension is lower than that of $\TT$ and they play a critical role in describing the usual measures of entanglement \cite{BCDLR}. This happens for instance for the Lee-Yang theory both at and away from criticality. The form factors and two-point functions of the branch point twist field and composite twist field for this theory were studied in \cite{BCD}. The expectation values of composite twist fields involving the energy field in the Ising  field theory were studied in \cite{ctheorem,Levi}.

\medskip
This paper is organised as follows: In Section 2 we review form factor results for the order and disorder fields in the Ising field theory as well as for $\TT$ and $\TT_\mu$. We review the cumulant expansion of two-point functions and introduce an example of the type of convergence issues that arise in the cumulant expansion of $\bra \TT_\mu(0)\TT^\dagger_\mu(r)\ket/{\bra \TT_\mu \ket^2}$. In Section 3 we find closed formulae for all higher cumulants, leading to a close expression for the two-point function. In Section 4 we test this expression by obtaining the exact conformal dimension of $\TT_\mu$ from resummation of leading terms in the short-distance expansion of the cumulants. We show that the normalised two-point function $\bra \TT_\mu(0)\TT_\mu^\dagger(r)\ket/{\bra \TT_\mu \ket^2}$ is in fact proportional to the normalised two-point function $\bra \mu(0)\mu(r)\ket/{\bra \mu \ket^2}$, thus it factorises into $n$-dependent and $n$-independent parts. In Section 5 we show how to analytically continue the cumulant expansion from $n$ integer and greater than 1 to $n$ real. This allows us to write a formula for the $n$-derivative of the two-point function at $n=1$, a quantity that typically plays a role in entanglement measures. We conclude in Section 6. Appendix A provides proofs of new useful resummation formulae for the form factors of $\TT_\mu$.

\section{Field Content of the Ising Model and Form Factors} 
 The correlation functions and form factors of the fields $\sigma, \mu$ defined by (\ref{simu}) can be obtained via form factor bootstrap \cite{KW,SmirnovBook} and  where studied in detail by Yurov and Zamolodchikov in their seminal paper \cite{YZam}.  Form factors of descendent fields (in the CFT sense) of the energy field $\varepsilon$ were studied in \cite{Cardybast} and shown to match in number and spin the field content of the corresponding Verma module in the underlying Ising CFT. 
 
Starting from the relations \eqref{simu} the form factor equations can be written and solved for matrix elements of $\sigma, \mu$ and these were found to take an extremely simple form \cite{YZam}, namely (the factor $i^k$ is needed to satisfy the kinematic residue equation):  
\beq
F_{2k}^{\mu}(\theta_1,\ldots,\theta_{2k})=i^k \bra \mu\ket \prod_{1\leq i<j \leq 2k} \tanh\frac{\theta_{ij}}{2}\,,
\eeq 
\beq
F_{2k+1}^{\sigma}(\theta_1,\ldots,\theta_{2k+1})=i^k F_1^\sigma \prod_{1\leq i<j \leq 2k+1} \tanh\frac{\theta_{ij}}{2}\,,
\eeq 
with $\theta_{ij}:=\theta_i-\theta_j$ and $\bra \mu \ket$ and $F_1^\sigma$ normalisation constants which can be identified with the vacuum expectation value of $\mu$ and the one-particle form factor of $\sigma$, respectively. More generally, a $k$-particle form factor is defined as
\beq
F_{k}^{\mathcal{O}}(\theta_1,\ldots,\theta_k):=\bra 0| \mathcal{O}(0)|\theta_1\ldots\theta_k|0 \ket\,
\eeq 
that is, a matrix element of a local or quasi-local field between the ground state $|0\ket$ and a $k$-particle state characterised by rapidities $\{\theta_i\}_k$. In general, particles will also be characterised by their quantum numbers, but in the Ising model there is a single particle type so these do not need to be specified. 

For the field $\mu$ the products above can be rewritten as a Pfaffian of a $2k \times 2k$ antisymmetric matrix $\mathcal{A}$ with entries ${\mathcal{A}}_{ij}=\tanh\frac{\theta_{ij}}{2}$. In particular this means that in the disordered phase the vacuum expectation value of $\mu$ is non-vanishing whereas it is vanishing for $\sigma$. 

In the context of the study of entanglement we know also that branch point twist fields $\TT$ and $\TT^\dagger$ play a prominent role. Their form factors in the (replica) Ising model have been known for some time \cite{entropy,nexttonext} and due to the free nature of the model they can also be expressed in terms of a Pfaffian
\beq
F_{2k}^{\TT|11\ldots 1}(\theta_1,\ldots,\theta_{2k};n)=\bra \TT\ket {\rm Pf}(K)\,,\qquad {\rm Pf}(K)=\sqrt{\det K}\,,
\eeq 
where $n$ labels the number of replicas, 
\beq
K_{ij}:=k(\theta_{ij})=\frac{\sin\frac{\pi}{n}}{2n \sinh\left(\frac{i\pi - \theta_{ij}}{2n}\right)\sinh\left(\frac{i\pi + \theta_{ij}}{2n}\right)}\frac{\sinh \frac{\theta_{ij}}{2n}}{\sinh \frac{i\pi}{2n}}\, \quad \mathrm{with} \quad i,j=1,\ldots, 2k\,,
\eeq 
and the superindices $11\ldots 1$ indicate that all particles are in the same copy 1. From this representation we also see that all form factors are functions of rapidity differences only, a property that holds for all spinless fields in relativistic quantum field theory. The two-particle form factor is simply $F_2^{\TT|11}(\theta_1,\theta_2;n)=\bra \TT \ket k(\theta_{12})$. Form factors for particles in copies $j_1\ldots j_{2k}$ can be obtained from the above using the standard form factor equations presented in \cite{entropy}
\beq 
F_{2k}^{\TT|j_1\ldots j_{2k}}(\theta_1,\ldots,\theta_{2k};n) = F_{2k}^{\TT|1\ldots 1} (\theta_1^{j_1-1},\ldots,\theta_{2k}^{j_k-1};n)\,, \qquad \mathrm{for}\qquad j_1\geq j_2\cdots\geq j_{2k}\,,
\label{ff index shift general}
\eeq 
with 
\beq
\theta^j:=\theta+2\pi i j\,.
\label{shift}
\eeq
 The form factors of the composite twist field $\TT_\mu$ where first obtained in \cite{SymResFF} and have again the Pfaffian structure typical of the Ising model, that is 
\beq
F_{2k}^{\TT_\mu|11\ldots 1}(\theta_1,\ldots,\theta_{2k};n)=  \bra \TT_\mu\ket{\rm Pf}(W)\,
\eeq 
with
\beq
\label{wfun}
W_{ij}:=w(\theta_{ij})=\frac{\sin\frac{\pi}{n}}{2n \sinh\left(\frac{i\pi - \theta_{ij}}{2n}\right)\sinh\left(\frac{i\pi + \theta_{ij}}{2n}\right)}\frac{\sinh\frac{\theta_{ij}}{n}}{\sinh\frac{i\pi}{n}}\,.
\eeq
 As we can see, this differs from $k(\theta)$ above only because $n$ is replaced by $n/2$ in the minimal part of the form factor (\textit{i.e.} the part that does not contain kinematic poles). However, this small change leads to some important differences, the main one being the asymptotic properties
\beq
\lim_{\theta \rightarrow \infty} k(\theta)=0 \qquad \mathrm{and} \qquad \lim_{\theta \rightarrow \pm \infty} w(\theta)=\pm \frac{i}{n}\,,
\label{asym1}
\eeq
as well as
\beq
\lim_{n\rightarrow 1} k(\theta)=0\qquad \mathrm{and} \qquad \lim_{n\rightarrow 1} w(\theta)=i \tanh\frac{\theta}{2}\,.
\label{intprop}
\eeq 
Note that the last equality simply shows that the two-particle form factor of $\TT_\mu$ reduces to that of $\mu$ for $n=1$, as expected. This extends to higher-particle form factors too. 
It is known from the study of many models and arguments such as those presented in \cite{DSC} that  the asymptotics of two particle form factors should be related to the value of a one-particle form factor. This is a consequence of so-called cluster decomposition in momentum space. In simple theories, as assumed in \cite{DSC}, this one-particle form factor would be that of the same field. However, in the Ising model, due to $\mathbb{Z}_2$ symmetry there is a mixing between form factors of $\mu$ and $\sigma$ and also those of $\TT_\sigma$ (defined as the composition of $\TT$ and $\sum_j \sigma_j$) and $\TT_\mu$ in such a way that:
\beq 
\lim_{\theta \rightarrow \pm\infty} w(\theta)=\pm{\tau^2}\,.
\label{tau2}
\eeq 
where $\tau:=F_1^{\TT_\sigma|1}$ is the one-particle form factor of $\TT_\sigma$, which by relativistic invariance is $\theta$ independent. Combining (\ref{tau2}) with (\ref{asym1}) we have that
\beq 
|F_1^{\TT_\sigma|1}|^2=|\tau|^2=\frac{1}{n}\,.
\eeq
Higher form factors of $\TT_\sigma$ can also be related to Pfaffians by employing a more general version of the cluster decomposition property. Namely
\beq 
\lim_{\theta_{2k+2}\rightarrow \infty} \bra \TT_\mu \ket^{-1}  F_{2k+2}^{\TT_\mu}(\theta_1,\ldots,\theta_{2k+2};n)= \tau F_{2k+1}^{\TT_\sigma}(\theta_1,\ldots,\theta_{2k+1};n).
\eeq 
Note that the prefactor $\bra \TT_\mu \ket^{-1}$ ensures that when $k=0$ both sides of the equation become $\tau^2$. In this way, the form factors $F_{2k+1}^{\TT_\sigma}(\theta_1,\ldots,\theta_{2k+1};n)$ can be computed systematically and it is easy to show that they can be written as a sum of Pfaffians involving $2k$ variables. In fact, we can show that 
\beq 
F_{2k+1}^{\TT_\sigma}(\theta_1,\ldots,\theta_{2k+1};n)=\frac{\tau}{\bra \TT_\mu \ket}  \sum_{j=1}^{2k+1} (-1)^{j+1} F_{2k}^{\TT_\mu}(\theta_1,\ldots,\bar{\theta}_j,\ldots \theta_{2k+1};n) \,,
\label{newfor}
\eeq 
where the sign  depends on the position of the variable $\theta_j$ and can be worked out by counting Wick contractions. Similarly, the symbol $\bar{\theta}_j$ means that this variable is removed, hence this is a sum over $2k$-particle form factors depending on a subset of the variables $\{\theta_1,\ldots,\theta_{2k+1}\}$. For instance 
\beqa 
F_{3}^{\TT_\sigma}(\theta_1,
\theta_{2},\theta_3;n)&=&\tau(w(\theta_{12})-w(\theta_{13})+w(\theta_{23}))\nonumber\\
&=&\frac{\tau}{\bra \TT_\mu\ket} (F_2^{\TT_\mu}(\theta_{1},\theta_2;n)-F_2^{\TT_\mu}(\theta_{1}, {\theta}_3;n)+F_2^{\TT_\mu}(\theta_2,{\theta}_3;n))\,,
\label{15}
\eeqa 
so that each ``contraction" $\theta_{ij}$ where $|i-j|$ is even produces one minus sign. The formula (\ref{newfor}) is, as far as we know, new and first presented here. However, this structure is the same as for the form factors of the field $\sigma$ which are obtained in the limit $n=1$, for which the function $w(\theta)$ reduces to a $\tanh$ (see Eq. (\ref{intprop})). The  special properties of the $\tanh$ function mean that formulae such as (\ref{newfor}) and (\ref{15}) can be shown to factorise also as products of  $\tanh$ functions.

In \cite{SymResFF} it was also shown that the form factor (\ref{wfun}) gives the correct conformal dimension of $\TT_\mu$ via the $\Delta$-sum rule \cite{DSC}. This dimension is \cite{kniz,dixon,Calabrese:2004eu,ctheorem,GS}
\beq
\Delta_{\TT_\mu}={\Delta_\TT}+\frac{\Delta_\mu}{n}=\frac{n}{48}+\frac{1}{24n} \qquad\mathrm{with} \qquad \Delta_\TT=\frac{1}{48}\left(n-\frac{1}{n}\right), \quad \Delta_\mu=\frac{1}{16}\,.
\label{dimension}
\eeq
In fact $\Delta_{\TT_\mu}=\Delta_{\TT_\sigma}$ since $\Delta_\mu=\Delta_\sigma$ even if, for symmetry reasons, $\Delta_\sigma$ cannot be obtained from the $\Delta$-sum rule. 

\subsection{Two-Point Function and Cluster Expansion}
In this paper we are interested in properties of the ground state two-point function of the field $\TT_\mu$.  In general we would like to write down an expansion of the form 
\beq
\label{leading}
\log \left(\frac{\langle \TT_\mu(0) \TT_\mu^\dagger(r) \rangle}{\langle \TT_\mu \rangle^2}\right)=\sum_{\ell=1}^{\infty} c_{\ell}^{\TT_\mu}(r;n) \stackrel{mr\ll 1}{\approx} -4\Delta_{\TT_\mu}\log (mr)-K_{\TT_\mu} \,,
\eeq
where the sum is over functions $c_{\ell}^{\TT_\mu}(r;n)$ known as cumulants,
 $\Delta_{\TT_\mu}$ is the conformal dimension of the field $\TT_\mu$ and $K_{\TT_\mu}$ is a constant that depends on the vacuum expectation value $\bra \TT_\mu \ket$. These cumulants are multiple integrals of linear combinations of products of form factors. More precisely, we have the following structure 
\beqa 
c^{\TT_\mu}_\ell(r;n)=\frac{1}{\ell!(2\pi)^{\ell} } \sum_{j_1,\ldots, j_{\ell}=1}^n \int_{-\infty}^\infty d\theta_1 \cdots \int_{-\infty}^\infty d\theta_{\ell} \, \, {h}^{\TT_\mu|j_1\ldots j_{\ell}}_{\ell}(\theta_1,\cdots,\theta_{\ell},n) e^{-mr \sum_{i=1}^{
\ell}\cosh\theta_i},
\label{genc}
\eeqa 
where the functions $h_k^{\mathcal{O}|j_1\ldots j_k}(\theta_1,\cdots, \theta_k,n)$ are given in terms of the form factors of the field involved, and $j_i$ represent the particle's quantum numbers which in our examples will be also the copy numbers. For example: 
\beqa 
h_2^{\TT_\mu|j_1 j_2}(\theta_{1},\theta_2,n)&=&\bra \TT_\mu \ket^{-2} \left|{F}^{\TT_\mu|j_1 j_2}_2(\theta_1,\theta_2,n)\right|^2, \nonumber\\
 h_4^{\TT_\mu|j_1 j_2 j_3 j_4}(\theta_{1},\theta_2,\theta_3, \theta_4, n)&=&\bra \TT_\mu \ket^{-2}\left|{F}^{\TT_\mu|j_1 j_2 j_3 j_4}_4(\theta_1,\theta_2,\theta_3,\theta_4, n)\right|^2\nonumber\\
&&
-h_2^{\TT_\mu|j_1 j_2}(\theta_{1},\theta_2,n)h_2^{\TT_\mu|j_3 j_4}(\theta_{3},\theta_4,n) \nonumber\\
 && -h_2^{\TT_\mu|j_1 j_3}(\theta_{1},\theta_3,n)h_2^{\TT_\mu|j_2 j_4}(\theta_{2},\theta_4,n)\nonumber\\ &&
 -h_2^{\TT_\mu|j_1 j_4}(\theta_{1},\theta_4,n)h_2^{\TT_\mu|j_2 j_3}(\theta_{2},\theta_3,n) , 
\label{exh}
\eeqa 
and so on, whereas all odd particle terms are vanishing.  Similar formulae can be written for the cumulants of $\TT_\sigma$ where only odd particle cumulants are non-vanishing.  A generic combinatorial/diagramatic construction of these functions can be found for instance in \cite{wight}. For a generic local field $\mathcal{O}$, it is standard to require
\beq
h_\ell^{\mathcal{O}|j_1\ldots j_\ell}(\theta_1,\cdots, \theta_\ell) \sim e^{-\theta_i} \qquad {\mathrm{for}} \qquad \theta_i \in \mathbb{R} \qquad \mathrm{and} \qquad \theta_i \rightarrow \infty\,.
\label{asi}
\eeq
Given the properties of the form factors presented in the previous section, we see that this property is not satisfied for the cumulants of the two-point function of $\TT_\mu$, or indeed for the cumulants of the two-point function of $\mu$ as shown in \cite{YZam}. In fact, the cumulant expansion is still convergent in both cases, but the leading behaviour for small $mr$ is harder to extract than in theories where (\ref{asi}) holds.  

\subsection{Two-Particle Contribution}
\label{conver}
The aim of this work is to find a general, compact form, for all terms in the expansion of $\bra \TT_{\mu}(0)\TT_{\mu}^\dagger(r)\ket/{\bra \TT_\mu\ket^2}$. One way to check the two-point function expansion is to recover the conformal dimension of the field by exact resummation of all terms which are proportional to $\log(mr)$ for $mr \ll 1$, that is the first term in (\ref{leading}).

Let us start by considering the simplest contribution to the connected part of the two-point function $\bra \TT_{\mu}(0)\TT_{\mu}^\dagger(r)\ket/{\bra \TT_\mu\ket^2}$, which has already  been studied in the literature \cite{SymResFF}.  The first non-vanishing contribution to the cumulant expansion comes from $h_2^{\TT_\mu|j_1 j_2}(\theta_{1},\theta_2,n)$, which is nothing but the normalised squared modulus of the two-particle form factor. Using \eqref{ff index shift general} the latter can be rewritten as
\beq 
\sum_{i,j=1}^n \left|F_2^{\TT_\mu|ij}(\theta_1,\theta_2)\right|^2=n \sum_{j=0}^{n-1} \left|F_2^{\TT_\mu|11}(\theta_1+2\pi i j,\theta_2)\right|^2=n \bra \TT_\mu \ket^2 \sum_{j=0}^{n-1} w((-\theta_{12})^j)w(\theta_{12}^{j})\,,
\eeq
where the superindex $j$ is defined as in (\ref{shift}). Thus we have
\beqa
c_2^{\TT_\mu}(r;n)&=&n \sum_{j=0}^{n-1} \int_{-\infty}^\infty  \int_{-\infty}^\infty \frac{d\theta_1 d\theta_2}{2(2\pi)^2} \, w((-\theta_{12})^j)w(\theta_{12}^{j})\, e^{-mr \cosh\theta_1-mr \cosh\theta_2} \nonumber\\
&=& \frac{n}{(2\pi)^2} \sum_{j=0}^{n-1} \int_{-\infty}^\infty  d\theta  \, w((-\theta)^{j})w(\theta^{j}) \,K_0(2mr\cosh\frac{\theta}{2})\,.
\label{integral}
\eeqa
The sum above has been computed in \cite{SymResFF} by using the cotangent trick and is given by
\beq
\label{sum_over-copies_identity_David}
\sum_{j=0}^{n-1}w((-\theta)^j)w(\theta^j)= -i \tanh\frac{\theta}{2} (w(2\theta+i\pi)+w(2\theta -i\pi))-\frac{1}{n}\,.
\eeq
This function tends asymptotically to the value $\frac{1}{n}$ for $|\theta|\rightarrow \infty$. This means that the usual procedure consisting of expanding the Bessel function for $mr \ll 1$ and isolating the $\log(mr)$ leading term, thus effectively removing the Bessel function from the integrand (\ref{integral}), now leads to a divergent integral. The integral is however not divergent, it simply needs to be done with care. We can rewrite \eqref{integral} as
\beqa
c_2^{\TT_\mu}(r;n)&=&
 \frac{n}{(2\pi)^2}  \int_{-\infty}^\infty  d\theta  \, \left[\sum_{j=0}^{n-1}w((-\theta)^{j})w(\theta^{j})-\frac{1}{n}\right] \,K_0(2mr\cosh\frac{\theta}{2})\nonumber\\
 &+&\frac{1}{(2\pi)^2} \int_{-\infty}^\infty  d\theta  \, K_0(2mr\cosh\frac{\theta}{2})\,.
\label{integral2}
\eeqa
In this form, the integral in the first line can be approximated for $mr\ll 1$ by expanding the Bessel function, giving a leading contribution which is proportional to $\log(mr)$. The integral in the second line can be computed exactly to
\beq
\int_{-\infty}^\infty  d\theta  \, K_0(2mr\cosh\frac{\theta}{2})=2K_0(mr)^2 \stackrel{mr\ll 1}{\approx}  -2(\log(mr))^2\,,
\eeq 
 so that, in this case, the leading small $mr$ contribution diverges as $(\log(mr))^2$ rather than $\log(mr)$. Thus, although the cumulant (\ref{integral}) is still well-defined, its leading small $mr$ behaviour is now dominated by $(\log(mr))^2$ instead of $\log(mr)$.  This is a consequence of the property (\ref{asi}) not holding in this case. Nonetheless, terms of order $(\log(mr))^2$ should cancel out when including further contributions in the form factor series as one expects to recover the $1/r^{4\Delta_{\TT_\mu}}$ behaviour of the two-point function at short distances. In the next sections we will show one particular way to recover the expected scaling (\ref{leading}) from our cumulant expansion. 

\section{Higher Particle Contributions: Closed Formulae}
Existing studies of the branch point twist field two-point function for free fermions \cite{nexttonext} and bosons \cite{FBoson} have revealed that the form of higher cumulants can be considerably simplified. This is because under sum over particle types and integration over the rapidities, many of the terms in the cumulant either cancel each other out or can be shown to be identical. In fact, it is possible to show that just as for the standard branch point twist field, and for the same reasons already discussed in \cite{nexttonext, FBoson} the cumulants of the two-point function of $\TT_\mu$ take the generic form 
\begin{align}
\label{cumulant_coefficients}
c_{2 \ell}^{\TT_\mu}(r;n) &= \frac{n}{2 \ell (2\pi)^{2 \ell}} \sum_{j_1,\dots,j_{2\ell - 1}=0}^{n-1}\left[\prod_{i=1}^{2\ell} \int_{-\infty}^{+\infty} \mathrm{d}\theta_i\,e^{-mr\cosh \theta_i}\right]\nonumber \\
&\times (-1)^\ell \left(w(\theta_{12}^{-j_1})\prod_{k=1}^{\ell-1}w(\theta_{2k+1\, 2k+2}^{j_{2k}-j_{2k+1}}))\right)\left(w(\theta_{1\,2\ell}^{j_{2\ell-1}})\prod_{k=1}^{\ell-1}w(\theta_{2k\, 2k+1}^{-j_{2k-1}+j_{2k}})\right)\,.
\end{align}
By using the fact that $w(\theta^{-j})=-w((-\theta)^j)$, we can change the sign of half of the factors in the second line, cancelling out the factor $(-1)^\ell$. The integrand becomes:
\beq
\label{fully_connected_1}
\sum_{j_1,\dots,j_{2\ell - 1}=0}^{n-1}
w((-\theta_{12})^j_1)w(\theta_{1\,2\ell}^{j_{2\ell-1}})\prod_{k=1}^{\ell-1}w(\theta_{2k+1\, 2k+2}^{j_{2k}-j_{2k+1}})w((-\theta_{2k\, 2k+1})^{j_{2k-1}-j_{2k}})\,.
\eeq
In order to evaluate the integrals (\ref{cumulant_coefficients}) it is convenient to  perform a change of variables whereby we first change the sign of all the even rapidities, without any change in the integration measure. Then, defining $\hat{\theta}_{ij} \equiv \theta_i + \theta_j$ the integrand becomes a function of rapidity \textit{sums} only:
\beq
\label{fully_connected_2}
\sum_{j_1,\dots,j_{2\ell - 1}=0}^{n-1}
w((-\hat{\theta}_{12})^{j_1})w(\hat{\theta}_{23}^{\,j_1 - j_2})w(\hat{\theta}_{34}^{\,j_2 - j_3})\dots w(\hat{\theta}_{2\ell-1\,2\ell}^{\,j_{2\ell - 2} - j_{2\ell - 1}})w(\hat{\theta}_{1\,2\ell}^{\,j_{2\ell-1}})\,.
\eeq
We will refer to this as a fully connected sum, meaning that all terms are cyclicly ``connected" both at the level of the rapidities and the summation indices. The sum (\ref{fully_connected_2}) and others of a similar type can be computed recursively as shown below and in Appendix A. 

\subsection{Recursive Formulae}
The sum (\ref{fully_connected_2}) can be carried out leading to generalisations of the following result
\beq
\label{sum_over-copies_identity_1}
f_1(x,y;n):=\sum_{j=0}^{n-1}w((-x)^j)w(y^j)= -\frac{i}{2}\frac{\sinh\left(\frac{x+y}{2}\right)}{\cosh \frac{x}{2}\cosh\frac{y}{2}}
[w(x+y+i\pi)+w(x+y -i\pi)]-\frac{1}{n}\,,
\eeq
which is presented here for the first time, although the case $x=y$ was obtained in \cite{SymResFF} and has already been reported in (\ref{sum_over-copies_identity_David}).
It is also useful to know that
\beq
\label{sum single w}
\sum_{j=0}^{n-1} w(x^j)=i \tanh\frac{x}{2}\,.
\eeq
A derivation of formulae (\ref{sum_over-copies_identity_1}), (\ref{sum single w}) and their generalisations to multiple sums (see below) is presented in Appendix A. 

For the branch point twist field of free fermions and bosons \cite{nexttonext,FBoson} a formula almost identical to (\ref{sum_over-copies_identity_1}) also holds, albeit without the term  $-\frac{1}{n}$. This term in fact makes the generalisation of this sum to multiple sums more complex for $\TT_\mu$ than for $\TT$. It can nonetheless be done as follows. Let us consider, as an example, the next sum in the series, namely a sum of the form
\beqa
\label{sum3}
&& \sum_{j_1,j_2=0}^{n-1}w((-x)^{j_1})w(y^{j_1-j_2})w(z^{j_2})=\sum_{j=0}^{n-1} f_1(x,y^{-j},n) w(z^{j})\,.
\eeqa
Repeated use of (\ref{sum_over-copies_identity_1}) and \eqref{sum single w} to simplify (\ref{sum3}) leads to
\beqa
\label{sum33}
&& \sum_{j_1,j_2=0}^{n-1}w((-x)^{j_1})w(y^{j_1-j_2})w(z^{j_2})=-\frac{i}{n}\left(\tanh\frac{x}{2}+\tanh\frac{y}{2}+\tanh\frac{z}{2}\right)\nonumber\\
&&+   \frac{1}{4}\frac{\cosh\left(\frac{x+y+z}{2}\right)}{\cosh \frac{x}{2}\cosh\frac{y}{2}\cosh\frac{z}{2}}\left[2w(x+y+z)+w(x+y+z+2i\pi)+w(x+y+z-2i\pi) \right]\,.
\eeqa
This special case gives a good indication of the kind of structures that emerge. We observe that the contribution in the second line has exactly the same structure as found for the branch point twist field in the free fermion theory \cite{nexttonext}. The terms in the first line form a symmetric polynomial on the variables $\tanh \frac{x}{2}, \tanh\frac{y}{2}, \tanh\frac{z}{2}$. The general structure for higher sums goes as follows: let us define
\beq
f_{\ell}(x_1,\ldots,x_{2\ell},n):=\frac{2i  (-1)^\ell \sinh\frac{x}{2}}{\prod_{i=1}^{2\ell}2\cosh\frac{x_i}{2}} \mathcal{F}_\ell(x;n) \,, \quad g_{\ell}(x_1,\ldots,x_{2\ell+1},n) :=\frac{2(-1)^{\ell+1} \cosh \frac{x}{2}}{\prod_{i=1}^{2\ell+1}2\cosh\frac{x_i}{2}} \mathcal{G}_\ell(x;n)\,
\label{fandg}
\eeq
where  $x:=\sum_i x_i$ in both cases, and
\beqa
\mathcal{F}_\ell(x;n) &:=&
\sum_{j=1}^\ell \binom{2\ell -1}{\ell - j}
\left[w(x^{j-\frac{1}{2}}) + w(x^{-j+\frac{1}{2}})\right]\,
\label{funf}\\
\mathcal{G}_\ell(x;n)&:=& \binom{2\ell }{\ell} w(x)+
\sum_{j=1}^\ell \binom{2\ell}{\ell - j}
\left[w(x^j) + w(x^{-j})\right]\,
\label{fung}
\eeqa 
with
\beq 
\lim_{|x| \rightarrow \infty} \mathcal{F}_\ell\left(x\,;n\right) ={ \mathrm{sgn}}(x) \frac{i}{n} 2^{2\ell-1}\,,\quad  
\lim_{|x| \rightarrow \infty} \mathcal{G}_\ell\left(x\,;n\right) ={ \mathrm{sgn}}(x) \frac{i}{n} 2^{2\ell}\,
\label{asym}
\eeq
and 
\beq 
 \mathcal{F}_\ell\left(x\,;1\right) =2^{2\ell-1} i\coth\frac{x}{2}  \,,\quad   \mathcal{G}_\ell\left(x\,;1\right) =2^{2\ell-1} i\tanh\frac{x}{2} \,.
\label{n141}
\eeq 
We can then compute the sum (\ref{fully_connected_2}) to
\beqa
&&\sum_{j_1,\dots, j_{2\ell-1}=0}^{n-1}w((-x_1)^j)w(x_2^{\,j_1 - j_2})\dots w(x_{2\ell-1}^{\,j_{2\ell - 2} - j_{2\ell - 1}})w(x_{2\ell}^{j_{2\ell-1}})\nonumber\\
&& \qquad =f_\ell(x_1,\ldots,x_{2\ell},n)
+ \frac{(-1)^\ell}{n} \sum_{j=0}^{\ell-1} \sigma_{2j}^{(2\ell)}\left(\tanh\frac{x_1}{2},\ldots,\tanh\frac{x_{2\ell}}{2}\right)\,,
\label{mainf}
\eeqa
whereas a similar sum involving an even number of indices can be evaluated to
\beqa
&&\sum_{j_1,\dots, j_{2\ell}=0}^{n-1}w((-x_1)^j)w(x_2^{\,j_1 - j_2})\dots w(x_{2\ell}^{j_{2\ell-1}-j_{2\ell}}) w(x_{2\ell+1}^{j_{2\ell}})\nonumber\\
&& \qquad =g_\ell(x_1,\ldots,x_{2\ell+1},n)
+ i \frac{(-1)^{\ell} }{n} \sum_{j=0}^{\ell-1} \sigma_{2j+1}^{(2\ell+1)}\left(\tanh\frac{x_1}{2},\ldots,\tanh\frac{x_{2\ell+1}}{2}\right)\,.
\label{mainfodd}
\eeqa
In both formulae, $\sigma_j^{(\ell)}(a_1,\ldots, a_\ell)$ is the elementary symmetric polynomial of order $j$ in $\ell$ variables, defined as
\beqa
\sigma_0^{(\ell)}(a_1,\ldots, a_\ell)=1 \qquad \mathrm{and} \qquad  
 \sigma_j^{(\ell)}(a_1,\ldots, a_\ell)=\sum_{1\leq i_1<i_2<\cdots <i_j \leq \ell} a_{i_1} a_{i_2}\cdots a_{i_j}\,.
 \eeqa
 These formulae can be proven by induction, similar to computations presented in \cite{nexttonext,FBoson}. The proofs are presented in Appendix A.

An interesting property of the formula (\ref{mainf}) and a consistency check of its validity is the fact that we must recover the cumulant expansion of $\bra \mu(0)\mu(r) \ket/{\bra \mu \ket^2}$  for $n=1$.
Indeed, from (\ref{n141}) it follows that
\beq
f_\ell(x_1,\ldots,x_{2\ell},1)=\frac{ (-1)^{\ell+1} \cosh\frac{x}{2}}{\prod_{i=1}^{2\ell}\cosh\frac{x_i}{2}}=(-1)^{\ell+1}\sum_{j=0}^{\ell} \sigma_{2j}^{(2\ell)}(\tanh\frac{x_1}{2},\ldots, \tanh\frac{x_{2\ell}}{2}) 
\,.
\label{46}
\eeq
Then, in the limit $n\rightarrow 1$ the only term remaining from the sum (\ref{mainf}) is the symmetric polynomial $\sigma_{2\ell}^{(2\ell)}(\tanh\frac{x_1}{2},\ldots, \tanh\frac{x_{2\ell}}{2})$ which is just the product of its arguments. This agrees exactly with the cumulant expansion of  $\log \bra \mu(0)\mu(r) \ket $ given in \cite{YZam}, formula (3.12a). Similarly, it can be shown that 
\beq
g_\ell(x_1,\ldots,x_{2\ell+1},1)=\frac{ i (-1)^{\ell+1} \sinh\frac{x}{2}}{\prod_{i=1}^{2\ell+1} \cosh\frac{x_i}{2}}=i(-1)^{\ell+1} \sum_{j=0}^{\ell} \sigma_{2j+1}^{(2\ell+1)}(\tanh\frac{x_1}{2},\ldots, \tanh\frac{x_{2\ell+1}}{2}) 
\,.
\label{471}
\eeq

\subsection{Main Result from this Section}
In summary, putting together the cumulant expansion (\ref{cumulant_coefficients}) with the sum formula (\ref{mainf}) we have the following exact formula for the logarithm of the correlation function of composite twist fields in the Ising model:
\beqa 
\log \left(\frac{\langle \TT_\mu(0) \TT_\mu^\dagger(r) \rangle}{\langle \TT_\mu \rangle^2}\right)&&=\sum_{\ell=1}^\infty c_{2\ell}^{\TT_\mu}(r;n)\nonumber\\
&& = \sum_{\ell=1}^\infty \frac{n}{2 \ell (2\pi)^{2 \ell}} \left[\prod_{i=1}^{2\ell} \int_{-\infty}^{+\infty} \mathrm{d}{\theta}_i\,e^{-mr\cosh {{\theta}}_i}\right]\left[ f_\ell(\hat{\theta}_{12},\ldots,\hat{\theta}_{2\ell-1\,2\ell},\hat{\theta}_{1\,2\ell},n)\right.\nonumber\\
&&   + \left. \frac{(-1)^\ell}{n} \sum_{j=0}^{\ell-1} \sigma_{2j}^{(2\ell)}\left(\tanh\frac{\hat{\theta}_{12}}{2},\ldots,\tanh\frac{\hat{\theta}_{2\ell-1\,2\ell}}{2}, \tanh\frac{\hat{\theta}_{1\,2\ell}}{2}\right) \right]\,.
\label{comp}
\eeqa 
We now proceed to check this expression for consistency by examining its leading short-distance behaviour. 

\section{Conformal Dimensions  from the Cumulant Expansion}
One possible way to test the cumulant expansion of the previous section is to obtain the  correct conformal dimension of the field $\TT_\mu$  by identifying the leading short-distance contributions to the sum over cumulants.  Note that this dimension was already recovered by $\Delta$-sum rule in \cite{SymResFF}, but the computation in that case only involved the two-particle form factor of $\TT_\mu$, whereas our study below involves all cumulants, thus providing a more extensive test of all the form factors and cumulants. 
Each cumulant is expected to contain a leading contribution which is proportional to $\log mr$ so that the overall sum gives (\ref{leading}) with dimension given by (\ref{dimension}). 
\medskip

First, let us return to our sum (\ref{fully_connected_2}) and change variables once more. We define
\beq
\label{x_variables}
x_i = \hat{\theta}_{i,i+1} \quad \text{for} \quad i=1,\dots,2\ell -1 \quad,\quad x_{2\ell}=\theta_{2\ell}\,,
\eeq
so that:
\beq
\label{x-theta_identities}
\theta_i = \sum_{j=i}^{2\ell}(-1)^{j-i}x_j \quad,\quad \sum_{i=1}^{2\ell}\theta_i = \sum_{i=1}^\ell x_{2i-1} \quad , \quad \hat{\theta}_{1,2\ell}=\sum_{i=1}^{2\ell -1}(-1)^{i-1}x_i\,.
\eeq
The Jacobian of the transformation from the $\theta$ variables to the $x$ variables is an upper triangular matrix with the diagonal terms being all $+1$, so the measure acquires no extra factor.
Applying this change of variables to \eqref{fully_connected_2} and expressing the result in the new variables \eqref{x_variables} we get:
\begin{align}
\label{summation_copy_indeces_final}
&\sum_{j_1,\dots, j_{2\ell-1}=0}^{n-1}w((-x_1)^j)w(x_2^{\,j_1 - j_2})\dots w(x_{2\ell-1}^{\,j_{2\ell - 2} - j_{2\ell - 1}})w((\sum_{i=1}^{2\ell -1}(-1)^{i-1}x_i)^{\,j_{2\ell-1}}) \nonumber \\ &= (-1)^\ell \frac{2i \sinh\left(\sum_{i=1}^{\ell}x_{2i-1}\right)}{2 \cosh\left(\frac{\sum_{i=1}^{2\ell -1}(-1)^{i-1}x_i}{2}\right)\prod_{i=1}^{2\ell-1}2\cosh\left(\frac{x_i}{2}\right)} \mathcal{F}_\ell\left(2\sum_{i=1}^\ell x_{2i-1}\,;n\right)\nonumber\\
& +\frac{(-1)^\ell}{n} \sum_{j=0}^{\ell-1} \sigma_{2j}^{(2\ell)}\left(\tanh\frac{x_1}{2},\ldots,\tanh\frac{x_{2\ell-1}}{2},\tanh\frac{\sum_{i=1}^{2\ell-1} (-1)^{i-1} x_{i}}{2} \right)\,,
\end{align}
with $\mathcal{F}_\ell(x;n)$ the function defined by (\ref{funf}), with the asymptotics (\ref{asym}). Recalling (\ref{46}) it is possible to also express the sum over symmetric polynomials in terms of products of hyperbolic functions as
\beqa 
&& \sum_{j=0}^{\ell-1} \sigma_{2j}^{(2\ell)}\left(\tanh\frac{x_1}{2},\ldots,\tanh\frac{x_{2\ell-1}}{2},\tanh\frac{\sum_{i=1}^{2\ell-1} (-1)^{i-1} x_{i}}{2} \right)\nonumber\\
&& \qquad = \frac{\cosh\left(\sum_{i=1}^{\ell}x_{2i-1}\right)}{\cosh\left(\frac{\sum_{i=1}^{2\ell -1}(-1)^{i-1}x_i}{2}\right)\prod_{i=1}^{2\ell-1}\cosh\left(\frac{x_i}{2}\right)}-\tanh\frac{\sum_{i=1}^{2\ell-1} (-1)^{i-1} x_{i}}{2} \prod_{i=1}^{2\ell-1} \tanh\frac{x_i}{2} \,.
\label{47}
\eeqa 
This rewriting will prove useful later on. 

\subsection{Exponential Factors}
Now let us look at the exponential factors in the integrand of (\ref{cumulant_coefficients}) and see what they look like in terms of the new variables $x_i$. From the first relation in \eqref{x-theta_identities} one has:
\beq
\sum_{j=i}^{2\ell-1}(-1)^{j-i}x_j = 
\begin{cases}
\theta_i - \theta_{2\ell}& \,\, \mathrm{for} \,\,i \,\, \text{even} \\
\theta_i + \theta_{2\ell}&  \,\, \mathrm{for}  \,\, i \,\, \text{odd} 
\end{cases}\,,
\eeq
so that
\begin{align}
& \sum_{i=1}^{2\ell}\cosh{\theta_i} = \cosh{\theta_{2\ell}} + \sum_{i=1}^{2\ell -1}\cosh(\theta_i - \theta_{2\ell} + \theta_{2\ell}) \nonumber \\ &= \cosh{\theta_{2\ell}} + 
\cosh{\theta_{2\ell}}\left(\sum_{i \, \text{even}} \cosh(\theta_i - \theta_{2\ell})+\sum_{i \, \text{odd}} \cosh(\theta_i + \theta_{2\ell})\right) \nonumber \\ 
& +\sinh{\theta_{2\ell}}\left(\sum_{i \, \text{even}} \sinh(\theta_i - \theta_{2\ell})-\sum_{i \, \text{odd}} \sinh(\theta_i + \theta_{2\ell})\right) \nonumber \\ 
&= \cosh x_{2\ell}\left[1+\sum_{i=1}^{2\ell-1}\cosh\left(\sum_{j=i}^{2\ell-1}(-1)^{j-i}x_j\right)\right] + \sinh x_{2\ell}\left[\sum_{i=1}^{2\ell-1}(-1)^i\sinh\left(\sum_{j=i}^{2\ell-1}(-1)^{j-i}x_j\right)\right]\,.
\end{align}
Therefore, since none of the functions in (\ref{summation_copy_indeces_final}) depends on $x_{2\ell}$ the integral on this variable can be carried out by making use of the identity \beq
\int_{-\infty}^{+\infty} \mathrm{d}t\,\exp(-A \cosh t - B \sinh t) = 2K_0\left(\sqrt{A^2 - B^2}\right)\,,
\eeq
giving 
\beq
\label{x_2l-1_integration}
\int_{-\infty}^{+\infty} \mathrm{d}x_{2\ell}\, e^{-mr\sum_{i=1}^{2\ell}\cosh{\theta_i}} = 2 K_0(mrd_{2\ell -1})\,,
\eeq
with
\beq
d_{2\ell - 1}^2 = \left[1+\sum_{i=1}^{2\ell-1}\cosh\left(\sum_{j=i}^{2\ell-1}(-1)^{j-i}x_j\right)\right]^2 - \left[\sum_{i=1}^{2\ell-1}(-1)^i\sinh\left(\sum_{j=i}^{2\ell-1}(-1)^{j-i}x_j\right)\right]^2\,.
\label{d2l1}
\eeq
The $mr \ll 1$ expansion of the modified Bessel function is:
\beq
K_0(mrd_{2\ell -1}) = -\log{mr} + \log{{2}} - \ln{d_{2\ell -1}} - \gamma + o(mrd_{2\ell -1})\,,
\label{koe}
\eeq 
from which the leading short distance contributions to the cumulant expansion can be obtained. It is worth mentioning that one  could also resum contributions proportional to the constant term $\log2 -\gamma$ in (\ref{koe}) and those should contribute to the $K_{\TT_\mu}$-term in (\ref{leading}), that is to the logarithm of $\bra \TT_\mu \ket$. A similar computation was carried out in \cite{FBoson} for $\bra \TT \ket$ in the free boson theory.

\subsection{Short-Distance Behaviour of the Cumulant Expansion}
Putting together \eqref{summation_copy_indeces_final}, (\ref{47}) and \eqref{x_2l-1_integration} in \eqref{cumulant_coefficients} we can split the cumulant into three contributions
\beq
c_{2 \ell}^{\TT_\mu}(r;n) ={c}^{(1)}_{2\ell}(r;n) + {c}^{(2)}_{2\ell}(r)+ {c}^{\mu}_{2\ell}(r)\,.
\label{53}
\eeq 
We will define these contributions as follows. First:
\begin{align}
{c}_{2 \ell}^{(1)}(r;n) &= \frac{2(-1)^\ell i n}{\ell(4\pi)^{2\ell}}\int_{-\infty}^{+\infty}\mathrm{d}x_1\,\dots \int_{-\infty}^{+\infty}\mathrm{d}x_{2\ell-1}\, K_0(mrd_{2\ell -1}) \nonumber \\  &\times \frac{\sinh\left(\sum_{i=1}^{\ell}x_{2i-1}\right)}{\cosh\left(\frac{\sum_{i=1}^{2\ell -1}(-1)^{i-1}x_i}{2}\right)\prod_{i=1}^{2\ell-1}\cosh\left(\frac{x_i}{2}\right)} \hat{\mathcal{F}}_\ell\left(2\sum_{i=1}^\ell x_{2i-1}\,;n\right)\,,\label{hatc2}
\end{align}
with 
\beq
\hat{\mathcal{F}}_\ell(x;n):=\mathcal{F}_\ell(x;n)-{\rm{sgn}}(x)\frac{i}{n} 2^{2\ell-1} \,.
\label{shift2}
\eeq 
This shift is motivated by the asymptotics (\ref{asym}) and ensures that the function $\hat{\mathcal{F}}_\ell(x;n)$ goes to zero for $|x|$ large. This in turn ensures the convergence of the integrals even when the Bessel function is approximated by its leading short-distance contribution $-\log(mr)$. 

The next contribution is then a combination of the first term in (\ref{47}) and the term introduced by the shift  (\ref{shift2}): \beqa
c_{2\ell}^{(2)}(r)&=& \frac{(-1)^\ell}{\ell(2\pi)^{2\ell}}\int_{-\infty}^{+\infty}\mathrm{d}x_1\,\dots \int_{-\infty}^{+\infty}\mathrm{d}x_{2\ell-1}\, K_0(mrd_{2\ell -1}) \nonumber\\
&& \times \left[ \frac{\cosh\left(\sum_{i=1}^{\ell}x_{2i-1}\right) - \sinh\left(\sum_{i=1}^{\ell}x_{2i-1}\right) {\rm{sgn}}(\sum_{i=1}^\ell x_{2i-1})  }{\cosh\left(\frac{\sum_{i=1}^{2\ell -1}(-1)^{i-1}x_i}{2}\right)\prod_{i=1}^{2\ell-1} \cosh\left(\frac{x_i}{2}\right)}\right]\label{s2l}\,.
\eeqa
Note that this contribution is $n$-independent.  Finally, the contribution ${c}^{\mu}_{2\ell}(r)$ is nothing but the cumulant of the expansion of $\bra \mu(0)\mu(r)\ket/{\bra \mu \ket^2}$ resulting from the last term (the product of $\tanh$ functions) in (\ref{47}):
\beqa
\label{c_mu_cumulant}
c_{2\ell}^{\mu}(r)&=& \frac{(-1)^{\ell+1}}{\ell(2\pi)^{2\ell}}\int_{-\infty}^{+\infty}\mathrm{d}x_1\,\dots \int_{-\infty}^{+\infty}\mathrm{d}x_{2\ell-1}\, K_0(mrd_{2\ell -1}) \nonumber\\
&& \qquad \times
{\tanh\left(\frac{\sum_{i=1}^{2\ell -1}(-1)^{i-1}x_i}{2}\right)\prod_{i=1}^{2\ell-1} \tanh\left(\frac{x_i}{2}\right)}\,.\label{cmu}
\eeqa
This may look a bit different from the cumulant presented in \cite{YZam} but this is simply due to the change of variables. Note also that in \cite{YZam} they implicitly take $\bra \mu \ket=1$ in the cumulant expansion. 
\subsection{Leading Contribution to ${c}_{2 \ell}^{(1)}(r;n) $}
In order to evaluate the integral (\ref{hatc2}) we can perform yet another (and final!) change of variables:
\beq
y = \sum_{i=1}^\ell x_{2i -1} \quad \Rightarrow \quad x_{2\ell - 1} = y - \sum_{i=1}^{\ell - 1}x_{2i-1}\quad , \quad \sum_{i=1}^{2\ell - 1}(-1)^{i-1}x_i = y - \sum_{i=1}^{\ell - 1}x_{2i}\,.
\eeq
so that, at short distances  $\sum_{\ell }{c}_{2 \ell}^{(1)}(r;n) \approx -z_n \log(mr)$ with
\begin{align}
{z}_n=&\sum_{\ell=1}^{\infty} \frac{(-1)^\ell 2 n i}{\ell (4\pi)^{2\ell}}\int_{-\infty}^{+\infty} \mathrm{d}y\, \sinh{y} \,\hat{\mathcal{F}}_\ell(2y; n) \int_{-\infty}^{+\infty}\mathrm{d}x_1\dots\int_{-\infty}^{+\infty}\mathrm{d}x_{2\ell - 2}\nonumber \\
& \left[\sech\left(\frac{y - \sum_{i=1}^{\ell - 1}x_{2i-1}}{2}\right)\prod_{i=1}^{\ell - 1}\sech\left(\frac{x_{2i-1}}{2}\right)\right]\left[\sech\left(\frac{y - \sum_{i=1}^{\ell - 1}x_{2i}}{2}\right)\prod_{i=1}^{\ell - 1}\sech\left(\frac{x_{2i}}{2}\right)\right] \nonumber \\
= &\sum_{\ell=1}^{\infty} \frac{(-1)^\ell 2 n i}{\ell (4\pi)^{2\ell}}\int_{-\infty}^{+\infty} \mathrm{d}y\, \sinh{y} \,\hat{\mathcal{F}}_\ell(2y; n) G_\ell^2(y)\,, 
\label{integ}
\end{align}
where, exactly as in \cite{nexttonext, FBoson}:
\begin{align}
G_\ell(y) = \int_{-\infty}^{+\infty}\mathrm{d}x_1\dots\int_{-\infty}^{+\infty}\mathrm{d}x_{\ell - 1}\left[\sech\left(\frac{y - \sum_{i=1}^{\ell - 1}x_i}{2}\right)\prod_{i=1}^{\ell - 1}\sech\left(\frac{x_i}{2}\right)\right] = \int_{-\infty}^{+\infty}\mathrm{d}a \frac{(2\pi)^{\ell -1}e^{i a y}}{\cosh^\ell{\pi a}}\,,
\end{align}
and the functions $G_\ell(y)$ can be evaluated explicitly to
\beq
G_{\ell}(y)=\frac{{(2\pi)}^{\ell-1}}{(\ell-1)!}\left\{\begin{array}{cc}
\frac{y}{\pi \sinh\frac{y}{2}} \prod_{j=1}^{\frac{\ell}{2}-1}(\frac{y^2}{\pi^2}+(2j)^2) & \rm{for}\,\, \ell \,\,\rm{even}\\
\frac{1}{\cosh\frac{y}{2}} \prod_{j=1}^{\frac{\ell-1}{2}}(\frac{y^2}{\pi^2}+(2j-1)^2) & \rm{for}\,\, \ell \,\,\rm{odd}
\end{array}\right.\,.
\eeq
By replacing $\mathcal{F}_\ell(x;n)$ by $\hat{\mathcal{F}}_\ell(x;n)$ in (\ref{hatc2}), we have ensured that the integrals (\ref{integ}) are convergent since $G_\ell(y) \sinh{y}$ is asymptotically polynomial in $y$ and $\hat{\mathcal{F}}_\ell(2y;n)$ is exponentially decaying. They can be evaluated with great precision and fitted to the function
\beq
z_n=\frac{1}{12}\left(n-\frac{1}{n}\right)+\frac{1}{4n}+ z'\,,
\label{formula}
\eeq 
with $z'=-0.217(4)$. This gives $4\Delta_{\TT_\mu}$ plus an additional constant $z'$ which should be cancelled by contributions coming from $c^{(2)}_{2\ell}(r)+c_{2\ell}^\mu(r)$. 

\begin{figure}[h!]
\begin{center}
	\includegraphics[width=7.5cm]{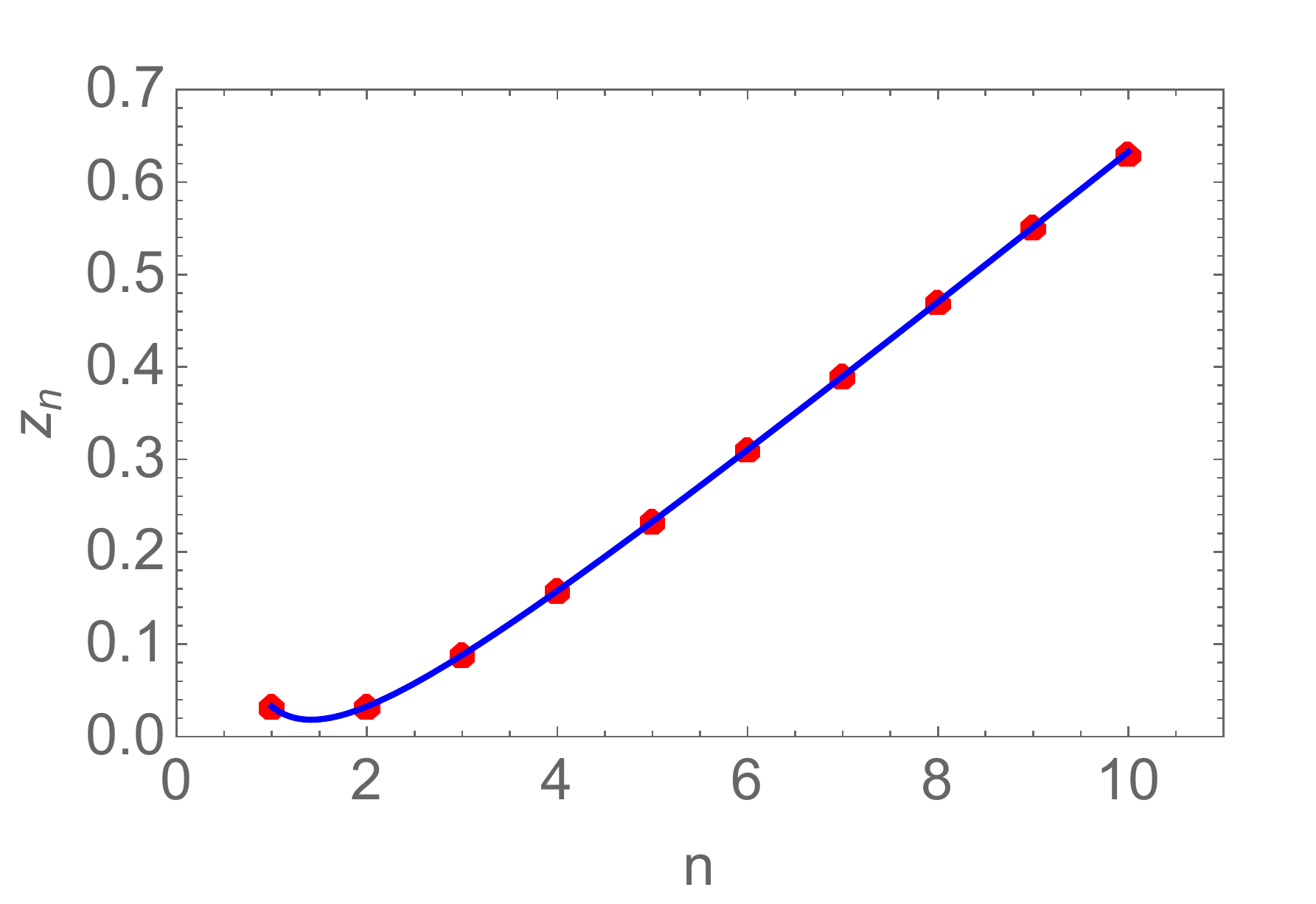}
		\includegraphics[width=7.5cm]{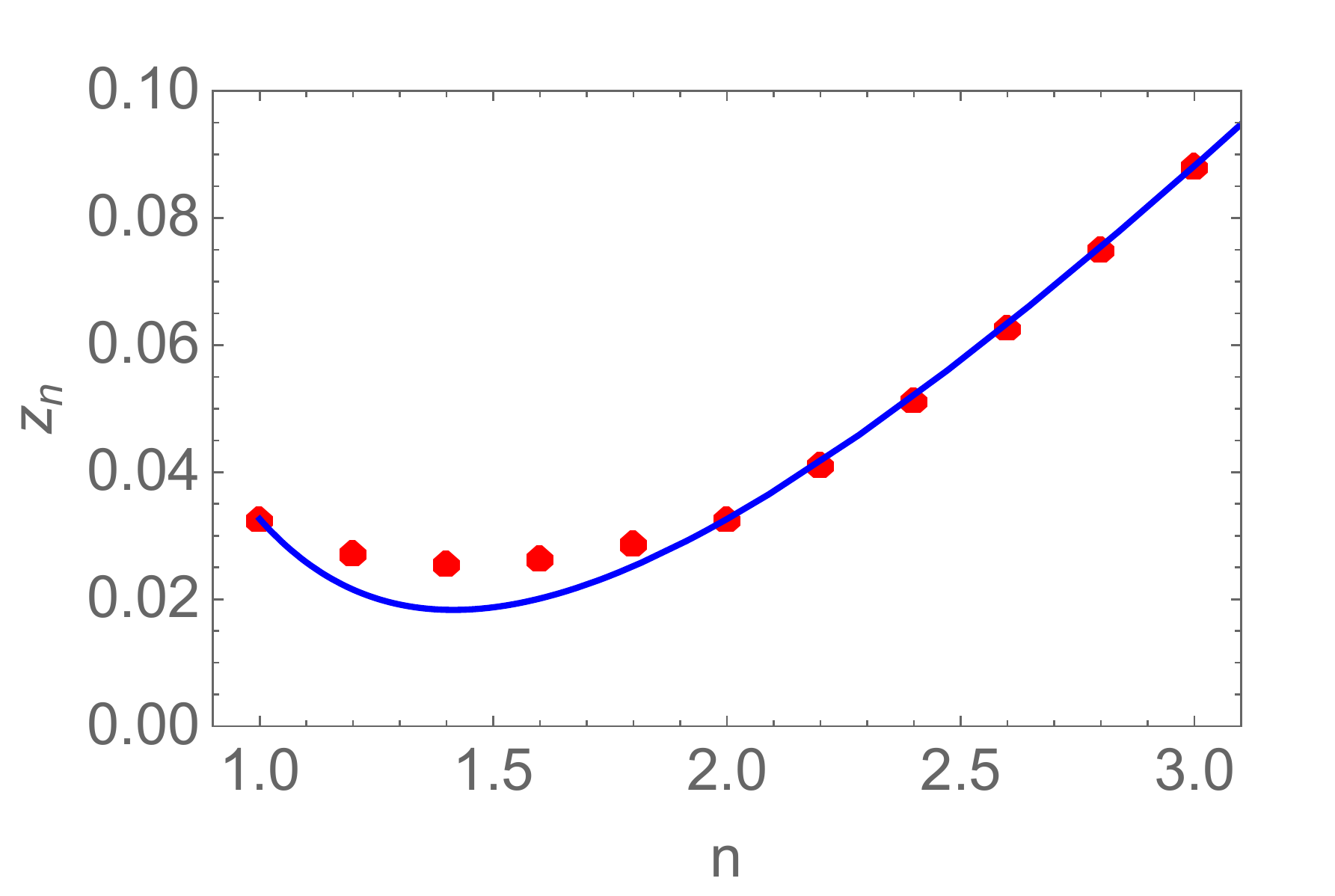}
				    \caption{Left: The function $z_n$ evaluated numerically through the sum (\ref{integ}) for integer values of $n=1,\dots ,10$ (red squares) against the formula (\ref{formula}) (blue solid line). Right: The same comparison for $n\in [1,3]$ including non integer values. When evaluating the sum \eqref{integ} numerically we truncate at some value of $\ell$. This value of $\ell$ is different for each value of $n$ and is chosen so that the sum is stable up to 5 decimal digits.}
				     \label{figure1}
    \end{center}
    \end{figure}

Numerical results for $z_n$ are shown in Fig.~\ref{figure1}.  It is interesting to observe that there is very good agreement with the formula (\ref{formula}) for $n$ integer and also for $n$ not integer, greater than 2. However for $1<n<2$ the numerical data differ from (\ref{formula}) suggesting that the analytic continuation of (\ref{integ}) to $n=1$ from $n$ real greater than 1 is non-trivial. This is in agreement with results found in \cite{SymResFF} where the limit $n\rightarrow 1$ of the two-particle form factor contribution produced a delta-function term. 
    
\subsection{Leading Contribution to ${c}_{2 \ell}^{(2)}(r)+c_{2\ell}^\mu(r)$}
Consider now the leading contribution to the second term in the cumulant. This is independent of $n$ and employing the same change of variables as above, it is easy to write an expression which is given by a convergent integral involving the functions $G_\ell(y)$. Letting $\sum_{\ell} {c}_{2 \ell}^{(2)}(r) \approx - z'' \log(mr)$ we have that
\begin{align}
z''= \sum_{\ell=1}^{\infty} \frac{2(-1)^\ell}{\ell (2\pi)^{2\ell}}\int_{0}^{+\infty} \mathrm{d}y\, e^{-y} \,G_\ell^2(y)=-0.0326(1)\,.
\label{integ2}
\end{align}
Note that 
\beq
z'+z''=-0.250(0)\approx -\frac{1}{4}\,.
\eeq
Remarkably, this value is precisely what we need to recover the correct dimension of the field $\TT_\mu$. This is because 
we know that 
\beq 
\sum_{\ell=1}^\infty c_{2\ell}^\mu(r)\approx -\frac{1}{4}\log(mr)\,,
\label{asymp}
\eeq 
as this is the sum over cumulants corresponding to the two-point function $\bra \mu(0)\mu(r)\ket/{\bra \mu \ket^2}$ and $\mu$ has dimension $4\Delta_\mu=1/4$. 
Therefore, the overall leading short distance behaviour of the $\bra \TT_\mu(0)\TT_\mu^\dagger(r) \ket/{\bra \TT_\mu \ket^2} $ cumulants correctly predicts the conformal dimension (\ref{dimension}). This highly non-trivial result provides strong support for the formula (\ref{comp}). In addition, the structure of the cumulants means that we can also write
\beq
\frac{\bra \TT_\mu(0)\TT_\mu^\dagger(r) \ket}{\bra \TT_\mu \ket^2}= {\mathcal{R}}(r;n)\, \frac{\bra \mu(0)\mu(r)\ket}{\bra \mu \ket^2}\,,
\eeq
where $\mathcal{R}(r;n)=\prod_{\ell=1}^\infty e^{{c}^{(1)}_{2\ell}(r;n)} e^{{c}^{(2)}_{2\ell}(r)}$ has the property $\mathcal{R}(r;1)=1$. 

Recalling the observation of Section \ref{conver}, namely that the   cumulant expansion of $\TT_\mu$ posed some convergence issues, we note that those issues did not feature in the computations of this section. This is because by writing the cumulant as we have done, all convergence issues have been ``hidden"  in the contribution $c_{2\ell}^\mu(r)$. Indeed, a naive expansion of the Bessel function in \eqref{c_mu_cumulant} leads to a divergent integral. Nonetheless, as shown in \cite{YZam}, the short distance limit of this quantity can be obtained via a semiclassical approach and it ultimately leads to the expected result (\ref{asymp}).

\section{Analytic Continuation to $n\in \mathbb{R}^{\geq 1}$} All results obtained so far are valid for $n\in \mathbb{Z}^+$. This is always the case in the replica picture where $n$ represents a replica number. However, the entanglement measures that our two-point function describes are typically defined for generic positive $n$. Therefore it is interesting to try and write an expression for the correlation function which is valid for $n\in \mathbb{R}^{\geq 1}$. Let us start by studying the analytic continuation of the leading short-distance terms.

\subsection{Analytic Continuation of Leading Short-Distance Contributions}
Fig.~\ref{figure1} (right) strongly suggests that our formula needs to be analytically continued in the region $1<n<2$. A similar problem was addressed in \cite{nexttonext,FBoson}, where is was shown that as $n$ approaches 1 from $n\gg 1$ some of the poles of the cumulants will cross or pinch the real line and provide additional contributions to the cumulant expansion which are non-vanishing for $n\in \mathbb{R}$ and need to be added. The correct analytic continuation is obtained when these contributions are correctly accounted for. The discussion is nearly identical as for the free boson case \cite{FBoson}, albeit involving different functions.

\medskip 
As we have seen, only the contribution $c_{2\ell}^{(1)}(r;n)$ to the cumulant is $n$-dependent. Therefore we only need to analytically continue the coefficient of the leading short-distance contribution to this term, that is $z_n$ defined in (\ref{integ}). For non-integer $n$ larger than 1, $z_n$ picks up additional contributions which account for the residues of the poles of $\hat{\mathcal{F}}_\ell(2y;n)$ that cross the real axis as $n \rightarrow 1^+$. The sum (\ref{funf}) in the function $\hat{\mathcal{F}}_\ell(2y,n)$ has kinematic poles at\footnote{The twist field approach assumes $n$ integer larger than 1 (since $n$ is a copy number). For that reason it is natural to look for an analytic continuation to $n=1$ from $n>1$. However, once found, the analytic continuation should be unique, thus valid for all $n$.} 
\beq 
2y \pm (2j-1)i\pi = (2k n + 1) i \pi \quad \text{and }\quad 2y \pm (2j-1)i\pi = (2k n - 1) i \pi \quad \text{for} \quad k\in \mathbb{Z}.
\eeq 
These poles result are due to the kinematic poles of the two-particle form factor (\ref{wfun}) at $\theta=i\pi$ and $\theta=i\pi(2n-1)$, together with those resulting from the periodicity property $w(\theta)=-w(-\theta+ 2\pi i n)$. This gives rise to four families of poles 
\beqa 
y_1&=&(kn+1-j)i\pi, \qquad y_2=(kn-j)i\pi,\qquad k \in \mathbb{Z}\\
y_3&=& (kn-1+j)i\pi,\qquad y_4=(kn+j)i\pi, \qquad k \in \mathbb{Z}, \label{genpoles}
\eeqa 
with corresponding residues of the function inside the sum (\ref{integ})  given by:
\beqa 
R_{1}(\ell,j,k,n)&=&\frac{n (-1)^{\ell+j}}{\ell(4\pi)^{2\ell}} \left(\begin{array}{c}
2\ell-1\\
\ell-j
\end{array}\right)\sinh(i\pi k n) G_\ell^2((nk-j+1)i\pi),\\
R_{2}(\ell,j,k,n)&=&-\frac{n (-1)^{\ell+j}}{\ell(4\pi)^{2\ell}} \left(\begin{array}{c}
2\ell-1\\
\ell-j
\end{array}\right)\sinh(i\pi k n) G_\ell^2((nk-j)i\pi),\\
R_{3}(\ell,j,k,n)&=&\frac{n (-1)^{\ell+j}}{\ell(4\pi)^{2\ell}} \left(\begin{array}{c}
2\ell-1\\
\ell-j
\end{array}\right)\sinh(i\pi k n) G_\ell^2((nk+j-1)i\pi),\\
R_{4}(\ell,j,k,n)&=&-\frac{n(-1)^{\ell+j}}{\ell(4\pi)^{2\ell}} \left(\begin{array}{c}
2\ell-1\\
\ell-j
\end{array}\right)\sinh(i\pi k n) G_\ell^2((nk+j)i\pi).
\eeqa 
These functions are all zero for $n$ integer but they  contribute for non-integer $n$. Let us now investigate which of these poles cross the real line in the limit $n\rightarrow 1^+$. 

Since there are many indices involved, let us start by considering just one example: $n=\frac{4}{3}$ and $\ell=3$ in the sum (\ref{integ}). According to the formula (\ref{dimension}) $4\Delta_{\TT_\mu}=0.236111$ in this case but the numerical evaluation of (\ref{integ}), after subtracting the constant $z'$, gives the value $0.243211$ which slightly overestimates the result. 
The disagreement is not simply due to numerical imprecision. The function $\hat{\mathcal{F}}_3(y,4/3)$ has poles that cross the integration line as $n\rightarrow 4/3$.
From (\ref{genpoles}) and the definition (\ref{funf}) we see that for $\ell=1$ the sum runs only over the value $j=1$. For $j=1$ the four families of poles labeled by the integer $k$ are:
\beqa 
y_1&=&ikn\pi, \qquad y_2=(kn-1)i\pi,\qquad k \in \mathbb{Z}\\
y_3&=& ikn\pi,\qquad y_4=(kn+1)i\pi, \qquad k \in \mathbb{Z}. \label{kpoles}
\eeqa 
 It is clear that all these poles are always above the real line (for $k>0$) or below the real line (for $k<0$), that is they never cross the real line, as $n$ approaches $\frac{4}{3}$. Therefore there is no correction coming from the $\ell=1$ contribution. Let us consider $\ell=2$. Now $j=1,2$. For $j=1$ the poles are the same as above and never cross the real line. For $j=2$ we have the following four families:
\beqa 
y_1&=&i(kn-1)\pi, \qquad y_2=(kn-2)i\pi,\qquad k \in \mathbb{Z}\\
y_3&=& i(kn+1)\pi,\qquad y_4=(kn+2)i\pi, \qquad k \in \mathbb{Z}. \label{kpoles2}
\eeqa 
We have already seen above that the poles $y_1$ and $y_3$ never cross the real line, so we can only have some contributions from $y_2$ and $y_4$.
For $k>0$ and $n$ positive and large both families of poles are above the real line. However, for $n=\frac{4}{3}$ we see that the pole $(kn-2)i\pi$ crosses the real line for $k=1$. Similarly, for $k<0$ and $n$ positive and large all poles are in the lower half plane but the pole $(kn+2)i\pi$ crosses the real line for $n=\frac{4}{3}$ and $k=-1$. 

\begin{figure}[h!]
\begin{center}
		\includegraphics[width=9cm]{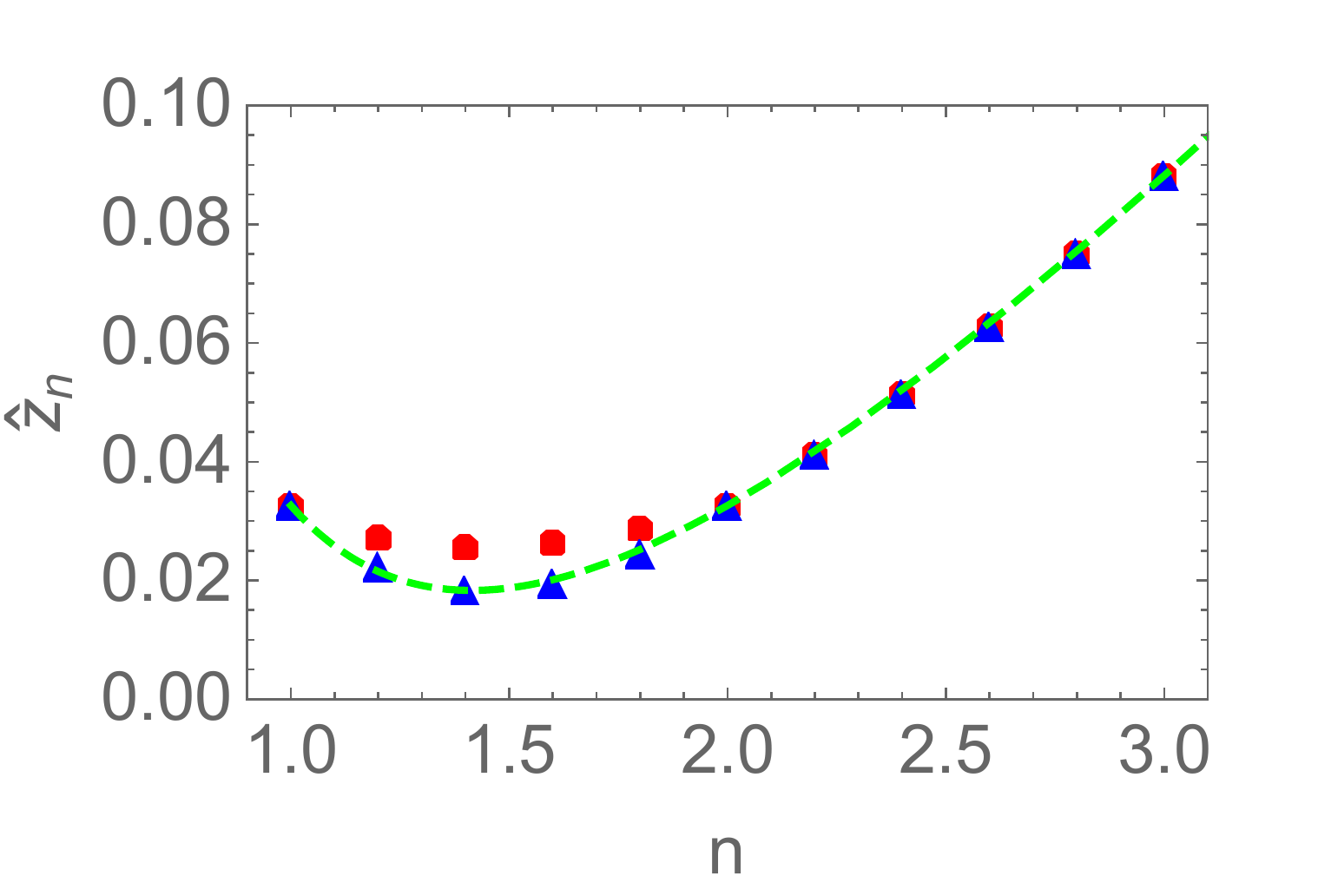}
				    \caption{The function $z_n$ evaluated numerically through the sum (\ref{integ}) for $n\in [1,3]$ (red squares) against the formula (\ref{formula}) (green dashed line) and its analytically continued values (blue triangles) given by (\ref{anul}).}
				     \label{figure2}
    \end{center}
    \end{figure}

In summary, there are two poles for $j=2$ located at $\pm \frac{2\pi i}{3}$.  The corresponding residue contributions are
\beqa  
 2\pi i (R_2(2,2,1,4/3)- R_4(2,2,-1,4/3))=-0.00680653\,. \label{exam32}
\eeqa  
Therefore, the addition of the residua of these two poles improves the estimate of the conformal dimension from $4\Delta_{\TT_\mu}= 0.243211$ to 
$4\Delta_\mu=0.243211-0.00680653=0.236404$ which is much closer to the exact value (note that the formula (\ref{integ}) gives -$4\Delta_{\TT_\mu}$, hence the minus sign of (\ref{exam32})). The addition of poles for higher values of $j$ will bring this value ever closer to formula (\ref{dimension}) as shown in Fig~\ref{figure2}.
In the general $n$ case, in order to fully identify those poles that will cross the real line we find once more four cases:
\beqa  
y_1: kn+1-j <0 \quad &\Rightarrow& \quad 1 \leq k < \frac{j-1}{n},\nonumber\\
y_2: kn-j<0 \quad &\Rightarrow& \quad 1 \leq k <\frac{j}{n},\nonumber\\
 y_3: kn-1+j <0 \quad &\Rightarrow & \quad -\frac{j-1}{n} < k \leq -1,\nonumber\\
 y_4: kn+j<0 \quad & \Rightarrow & \quad -\frac{j}{n} < k \leq -1,\label{ranges}
\eeqa  
This gives the analytically continued values $\hat{z}_n$ 
\beqa 
\hat{z}_n &=&z_n+\sum_{\ell=1}^\infty\sum_{j=1}^\ell \sum_{k=1}^{[\frac{j-1}{n}]-q_1} \frac{i n (-1)^{\ell+j+1}}{\ell (4\pi)^{2\ell-1}} \left(\begin{array}{c}
2\ell-1\\
\ell-j
\end{array}\right) \sinh\left(i\pi n k\right) G_\ell^2\left(\left(nk-j+1\right)i\pi\right)\nonumber\\
&&\quad +\sum_{\ell=1}^\infty\sum_{j=1}^\ell \sum_{k=1}^{[\frac{j}{n}]-q_2} \frac{i n (-1)^{\ell+j+1}}{\ell (4\pi)^{2\ell-1}} \left(\begin{array}{c}
2\ell-1\\
\ell-j
\end{array}\right) \sinh\left(i\pi n k\right) G_\ell^2\left(\left(nk-j\right)i\pi\right)\,, \label{anul}
\eeqa 
where we used the fact that the residues $R_2(\ell,j,k,n)=-R_4(\ell,j,k,n)$ and $R_1(\ell,j,k,n)=-R_3(\ell,j,k,n)$ (which produces a factor 2) and multiplied by $2\pi i$ as required by the residue theorem. 
The shifts $q_1, q_2$ take the value 1 when $n[\frac{j-1}{n}]=j-1$ and $n[\frac{j}{n}]=j$, respectively and are zero otherwise (they can be removed by requiring $n$ to be non-integer). Here the symbol $[.]$ represents the integer part. Fig.~\ref{figure2} shows the same functions as in Fig.~\ref{figure1} (right) plus an additional set of values, which are the analytically continued values of $z_n$ (in blue). As we can see these now agree perfectly with the fit (\ref{formula}), even for non-integer $n$ between 1 and 2. 
\subsection{Analytic Continuation of the $n$-Derivative}
Applications of the correlation function (\ref{comp}) in the context of entanglement measures frequently requires the computation of its derivative with respect to $n$ followed by the limit $n\rightarrow 1$. As discussed in \cite{entropy,nexttonext} and \cite{SymResFF} the derivative with respect to $n$ of the function (\ref{fandg}) has a discontinuity. More precisely, as $n$ approaches 1 and poles cross the real line, the derivative is not uniformly convergent as a function of $\theta$ and this leads to terms involving $\delta$-functions. The simplest examples of this phenomenon are seen for the two-particle contribution to the two-point function of $\TT$ \cite{entropy} and of $\TT_\mu$ \cite{SymResFF}. Here we show how this generalises to the whole cumulant sum. Notice that we only need to consider the contribution from the function $c_{2\ell}^{(1)}(r;n)$ in (\ref{hatc2}) since all other terms are independent of $n$ and so the derivative is zero. For this term, we actually only need to consider $\mathcal{F}(x;n)$ as the additional term in the ``hatted" version is also $n$-independent. So, we define
\beqa
s_{2\ell}^{\TT_\mu}(r)&:=&-
\lim_{n\rightarrow 1} \frac{d}{dn} c_{2\ell}^{(1)}(r;n)\nonumber\\
&=&\frac{2(-1)^{\ell+1} i }{\ell(4\pi)^{2\ell}}\int_{-\infty}^{+\infty}\mathrm{d}x_1\,\dots \int_{-\infty}^{+\infty}\mathrm{d}x_{2\ell-1}\, K_0(mrd_{2\ell -1}) \nonumber \\  
&&\times \frac{\sinh\left(\sum_{i=1}^{\ell}x_{2i-1}\right)}{\cosh\left(\frac{\sum_{i=1}^{2\ell -1}(-1)^{i-1}x_i}{2}\right)\prod_{i=1}^{2\ell-1}\cosh\left(\frac{x_i}{2}\right)} \lim_{n\rightarrow 1} \frac{d}{dn} \left[n {\mathcal{F}}_\ell\left(2\sum_{i=1}^\ell x_{2i-1}\,;n\right)\right]\,.
\eeqa
 One way to treat the derivative is to recall the $\ell=1$ result that was derived in \cite{SymResFF}, namely
\beqa
\lim_{n\rightarrow 1} \frac{d}{dn} n f_1(x,x,n)&=&
-\frac{i}{2} \frac{\sinh x}{\cosh^2 \frac{x}{2}}\lim_{n\rightarrow 1}\frac{d}{dn} n [w(2x+ i\pi) + w(2x- i\pi)]\nonumber\\
&=&\frac{x}{\cosh^2 \frac{x}{2} \sinh x}-\frac{\pi^2}{2} \delta(x)\,,
\label{ipishift}
\eeqa 
that is, there is a finite part and a distribution part that accounts for the behaviour around $x=0$. Recall that the function $f_1(x,y,n)$ is defined in (\ref{sum_over-copies_identity_1}). This extends to higher cumulants in similar ways, so that we can write
\beq
s_{2\ell}^{\TT_\mu}(r)=s_{2\ell}^{{\rm fin}}(r) + s_{2\ell}^{\delta}(r)\,,
\eeq 
where the two contributions represent the ``finite" and $\delta$-function contributions. The finite part can be easily computed by noting that
\beq
\lim_{n\rightarrow 1} \frac{d}{dn} n \sinh x \mathcal{F}(2x;n)=\frac{i 2^{2\ell-1} x}
{\sinh x}\,,
\eeq 
Therefore
\beqa
s_{2\ell}^{{\rm fin}}(r)
&=&\frac{(-1)^{\ell} }{\ell(2\pi)^{2\ell}}\int_{-\infty}^{+\infty}\mathrm{d}x_1\,\dots \int_{-\infty}^{+\infty}\mathrm{d}x_{2\ell-1}\, K_0(mrd_{2\ell -1}) \nonumber \\  
&&\times \frac{\sum_{i=1}^{\ell}x_{2i-1} }{\sinh\left(\sum_{i=1}^{\ell}x_{2i-1}\right)\cosh\left(\frac{\sum_{i=1}^{2\ell -1}(-1)^{i-1}x_i}{2}\right)\prod_{i=1}^{2\ell-1}\cosh\left(\frac{x_i}{2}\right)}\,.
\eeqa
The $\delta$-function contribution is a generalisation of the $\ell=1$ case seen above and can be obtained by identical arguments as those presented in \cite{nexttonext}. In fact, the result is also identical to formula (4.6) in \cite{nexttonext}, that is, 
\beqa 
s_{2\ell}^\delta(r)&=&\frac{\pi^2 (-1)^\ell}{\ell (4\pi)^{2\ell}}\int_{-\infty}^{+\infty}\mathrm{d}x_1\,\dots \int_{-\infty}^{+\infty}\mathrm{d}x_{2\ell-1}\, \delta(\sum_{i=1}^\ell x_{2i-1})\nonumber\\
&&\times \left[\left(\begin{array}{c}
2\ell-2\\
\ell-1
\end{array}\right)\frac{2K_0(2mrd_{2\ell-1})}{\cosh\left(\frac{\sum_{i=1}^{2\ell -1}(-1)^{i-1}x_i}{2}\right)\prod_{i=1}^{2\ell-1}\cosh\left(\frac{x_i}{2}\right)}  \right]\nonumber\\
&& - \frac{\pi^2 (-1)^\ell}{\ell (4\pi)^{2\ell}}\int_{-\infty}^{+\infty}\mathrm{d}x_1\,\dots \int_{-\infty}^{+\infty}\mathrm{d}x_{2\ell}\, \delta(\sum_{i=1}^\ell x_{2i-1})\nonumber\\
&&\times \sum_{j=1}^\ell \sum_{k=1}^{j-1} \sum_{q=\pm} \left[\left(\begin{array}{c}
2\ell-1\\
\ell-j
\end{array}\right) (-1)^j\frac{\prod_{i=1}^{2\ell} e^{-rm \cosh\left(\sum_{j=1}^{2\ell}(-1)^{j-i}x_i+i\pi q \frac{j-k}{2\ell}\right)}}{\cosh\left(\frac{\sum_{i=1}^{2\ell -1}(-1)^{i-1}x_i}{2}\right)\prod_{i=1}^{2\ell-1}\cosh\left(\frac{x_i}{2}\right)}  \right]\,.
\eeqa 

\section{Conclusion}
In this paper we have studied the normalised two-point function $\bra \TT_\mu(0)\TT^\dagger_\mu(r) \ket/{\bra \TT_\mu \ket^2}$ of the composite twist field $\TT_\mu$ and its conjugate. The motivation to study this object comes from recent investigations of a measure of entanglement known as symmetry resolved entanglement \cite{GS,german3,SymResFF}. More fundamentally, our work contributes to developing the understanding of correlation functions in the replica Ising field theory, a theory that, although free and seemingly simple, contains a large number of symmetry fields or twist fields which are not present in the standard, non-replicated, model. 

The current work uses traditional IQFT techniques, mainly the form factor bootstrap program adapted to composite twist fields \cite{SymResFF}, to expand the logarithm of the correlation function into a series of cumulants. The main result of the paper is finding simplified expressions for these cumulants which result from proving a number of multiple sum formulae, presented in  Appendix A, involving the two-particle form factors of the field $\TT_\mu$. 

Employing this cumulant expansion we have found the following structure 
\beq 
\frac{\bra \TT_\mu(0)\TT^\dagger_\mu(r) \ket}{\bra \mu(0)\mu(r) \ket} \frac{\bra \mu\ket^2}{\bra \TT_\mu\ket^2}=\mathcal{R}(r;n) \, \qquad \mathrm{with} \qquad \mathcal{R}(r;1)=1\,,
\label{factor}
\eeq 
where $\mu$ is the disorder field of the Ising field theory, and $\mathcal{R}(r;n)$ has an explicit form given in terms of a multiple integral representations which we describe in detail in this paper. By exact resummation of leading contributions to the cumulant expansion, we have shown that at short distances this two-point function scales as a power law in $r$ with the power (\ref{dimension}) which is exactly predicted by CFT. Showing that the two-point correlation function of the field $\mu$ factors out of the correlator (\ref{factor}) has been a crucial ingredient in recovering the correct scaling dimension. We have also shown how our formulae may be analytically continued to real replica number and how non-analytic, delta-function terms emerge for all cumulants when computing the derivative w.r.t. to $n$. This generalises results found in \cite{SymResFF} and \cite{entropy}. 

As a byproduct of our investigation, we have also shown that the form factors of the composite field $\TT_\sigma$ can be obtained from those of $\TT_\mu$ via clustering in momentum space, in much the same way as the form factors of the fields $\sigma$ and $\mu$ are related. We have also fixed the normalisation of the $\TT_\sigma$ form factors by fixing the one-particle form factor from the asymptotics of the two-particle form factor of $\TT_\mu$.

As mentioned above, our result has applications in the context of the symmetry resolved entanglement entropy and directly leads to a more complete formula for the latter in the Ising model. We further expect the results of this investigation to apply with some modifications to other composite fields, at least for free theories, for instance those associated with $U(1)$ symmetry in doubled free models which were studied in \cite{Bonsignori_2019, Murciano_2020, Horvath_2021}. 

\medskip {\bf Acknowledgments:} The authors thank Benjamin Doyon and D\'avid X. Horv\'ath for useful discussions. Michele Mazzoni is grateful for funding under the EPSRC Mathematical Sciences Doctoral Training Partnership EP/W524104/1. Olalla A. Castro-Alvaredo thanks EPSRC for financial support under Small Grant EP/W007045/1 and the Kavli Institute for Theoretical Physics (Santa Barbara) for financial support
from the National Science Foundation under Grant No. NSF PHY-1748958,
and hospitality during the conference ``Talking Integrability: Spins, Fields and  Strings", August 29-September 1 (2022).

\appendix
\section{Summation formulae}
Let us first prove the identities \eqref{sum single w} and \eqref{sum_over-copies_identity_1}, which are both obtained via the cotangent trick. To show \eqref{sum single w}, consider the integral:
\beq
\frac{1}{2\pi i}\oint_\mathcal{C}\mathrm{d}z\, \pi \cot(\pi z)w(x^z)
\eeq
where $\mathcal{C}$ is the rectangular contour with vertices $-\epsilon + i L$, $-\epsilon - i L$, $n-\epsilon - i L$, $n-\epsilon + i L$. The vertical contributions cancel off because the integrand is invariant under the shift $z \to z + n$. The same holds for the horizontal contributions in the large $L$ limit, as
\beq
\lim_{L \to \infty} \cot{\pi(t\pm iL)} = \mp i \,,\quad \lim_{L \to \infty} w\left(x^{t\pm i L}\right)= \mp \frac{i}{n}
\eeq
and therefore the sum of the residues must vanish. Within the integration contour, the function $\pi \cot(\pi z)$ has simple poles at $z=0,\,1,\dots\,n-1$ with unite residue. The kinematic poles of $w(x^z)$ are at $z= \frac{1}{2}- \frac{x}{2\pi i}$, $z= n -\frac{1}{2}- \frac{x}{2\pi i}$, with residue $\frac{1}{2\pi}$. At both these points, $\cot(\pi z) = -i \tanh{\frac{x}{2}}$. Putting all the pieces together, one has therefore:
\beq 
0=\sum_{j=0}^{n-1}w(x^j) - i \tanh{\frac{x}{2}}\,.
\eeq
Using the very same strategy, one can prove \eqref{sum_over-copies_identity_1}. The integral to evaluate is now
\beq
\frac{1}{2\pi i} \oint_\mathcal{C} \mathrm{d}z\,\pi\cot(\pi z)w((-x)^z)w(y^z)\,,
\eeq
along the same contour $\mathcal{C}$ as before. In this case, however, the horizontal contributions do not cancel off, as
\beq
\lim_{L \to \infty} w((-x)^{t \pm iL}))w(y^{t \pm iL}) = -\frac{1}{n^2}\,,
\eeq
and thus the integral evaluates to $-\frac{1}{n}$ in the large $L$ limit. Summing over the residues of the poles of $\pi \cot{(\pi z)}$ gives the left-hand side of \eqref{sum_over-copies_identity_1}, while the kinematic poles are now at $z = \frac{1}{2}+\frac{x}{2\pi i}\,,\,-\frac{1}{2}+n + \frac{x}{2\pi i}$ and $z = \frac{1}{2}-\frac{y}{2\pi i}\,,\,-\frac{1}{2}+n - \frac{y}{2\pi i}$, with residues:
\begin{align*}
&\underset{z = \frac{1}{2}+\frac{x}{2\pi i}}{\text{Res}}w((-x)^z)w(y^z) = \frac{1}{2\pi}w(x+y + i\pi)\,, \quad &\underset{z = n-\frac{1}{2}+\frac{x}{2\pi i}}{\text{Res}}w((-x)^z)w(y^z) = \frac{1}{2\pi}w(x+y - i\pi) \\
&\underset{z = \frac{1}{2}-\frac{y}{2\pi i}}{\text{Res}}w((-x)^z)w(y^z) = -\frac{1}{2\pi}w(x+y - i\pi)\,, \quad &\underset{z = n-\frac{1}{2}-\frac{y}{2\pi i}}{\text{Res}}w((-x)^z)w(y^z) = -\frac{1}{2\pi}w(x+y + i\pi)\,.
\end{align*}
Evaluating the cotangent at the kinematic poles and putting all the pieces together, one has
\begin{align*}
-\frac{1}{n} &=\sum \text{Res}[\pi \cot(\pi z) w((-x)^z)w(y^z) \\&= \sum_{j=0}^{n-1}w((-x)^j)w(y^j) + \frac{i}{2}\frac{\sinh\left(\frac{x+y}{2}\right)}{\cosh\left(\frac{x}{2}\right)\cosh\left(\frac{y}{2}\right)}
[w(x+y+i\pi)+w(x+y -i\pi)]\,,
\end{align*}
which is indeed \eqref{sum_over-copies_identity_1}.

Using this result, it is possible to prove \eqref{mainf} and \eqref{mainfodd} by induction. It is useful to observe beforehand that the following expansions hold:
\begin{align}
\label{trig identities}
\frac{\sinh{\left(\sum_{i=1}^{k} x_i\right)}}{\prod_{i=1}^{k} \cosh{x_i}} &= \sum_{j=0}^{[\frac{k-1}{2}]} \sigma_{2j+1}^{(k)}(\tanh{x_1},\dots,\tanh{x_{k}}) \\
\frac{\cosh{\left(\sum_{i=1}^{k} x_i\right)}}{\prod_{i=1}^{k} \cosh{x_i}} &= \sum_{j=0}^{[\frac{k}{2}]} \sigma_{2j}^{(k)}(\tanh{x_1},\dots,\tanh{x_{k}})\,.
\end{align}
Following the procedure employed in  \cite{nexttonext,FBoson} we will prove that \eqref{mainf} implies \eqref{mainfodd}. If \eqref{mainf} holds, than we can shift $x_{2\ell}$ by $-2i\pi p$, multiply the left-hand side by a factor $w(x_{2\ell + 1}^p)$ and sum over $p$ to obtain:
\begin{align}
\label{induction proof}
&\sum_{j_1,\dots, j_{2\ell-1},p=0}^{n-1}w((-x_1)^j)w(x_2^{\,j_1 - j_2})\dots w(x_{2\ell-1}^{\,j_{2\ell - 2} - j_{2\ell - 1}})w(x_{2\ell}^{j_{2\ell-1}-p})w(x_{2\ell + 1}^p) \nonumber\\
&=\sum_{p=0}^{n-1}f_\ell(x_1,\ldots,x_{2\ell}^{-p},n)w(x_{2\ell + 1}^p)
+ \sum_{p=0}^{n-1}\frac{(-1)^\ell}{n} \sum_{j=0}^{\ell-1} \sigma_{2j}^{(2\ell)}\left(\tanh\frac{x_1}{2},\ldots,\tanh\frac{x_{2\ell}^{-p}}{2}\right)w(x_{2\ell + 1}^p)\,,
\end{align}
Let us focus on the first term in the second line, which yields two contributions due to the presence of a constant term in the right-hand side of \eqref{sum_over-copies_identity_1}. Defining $x = \sum_{i=1}^{2\ell}x_i$, this reads:
\begin{align}
\label{app first term}
&\frac{2i  (-1)^\ell \sinh\frac{x}{2}}{\prod_{i=1}^{2\ell}2\cosh\frac{x_i}{2}}
\sum_{p=0}^{n-1}\sum_{j=1}^\ell \binom{2\ell -1}{\ell - j}
\left[w\left(x^{j-p} - i\pi \right) + w\left(x^{-j-p} + i\pi\right)\right]w(x_{2\ell + 1}^p) \nonumber \\
&=g_\ell(x_1,\dots,x_{2\ell +1};n) + \frac{4i  (-1)^\ell \sinh\frac{x}{2}}{n\prod_{i=1}^{2\ell}2\cosh\frac{x_i}{2}}\sum_{j=1}^\ell \binom{2\ell -1}{\ell - j} \nonumber \\
&=g_\ell(x_1,\dots,x_{2\ell +1};n) + \frac{i(-1)^\ell}{n}\sum_{j=0}^{\ell -1} \sigma_{2j+1}^{(2\ell)}(\tanh{\frac{x_1}{2}},\dots,\tanh{\frac{x_{2\ell}}{2}})
\end{align}
The emergence of the function $g_\ell$ was already proved in Appendix C of  \cite{FBoson}. In going from the second to the third line we used the first identity in \eqref{trig identities} and $\sum_{j=1}^\ell \binom{2\ell -1}{\ell - j} = 2^{2\ell -2}$. Consider now the second term in the second line of \eqref{induction proof}: using the fact that $\tanh{\frac{x^{\pm p}}{2}}=\tanh{\frac{x}{2}}$ and \eqref{sum single w} we have:
\begin{align}
\label{app second term}
& \sum_{p=0}^{n-1}\frac{(-1)^\ell}{n} \sum_{j=0}^{\ell-1} \sigma_{2j}^{(2\ell)}\left(\tanh\frac{x_1}{2},\ldots,\tanh\frac{x_{2\ell}^{-p}}{2}\right)w(x_{2\ell + 1}^p) \nonumber \\
&=\frac{i(-1)^\ell}{n} \sum_{j=0}^{\ell-1} \sigma_{2j}^{(2\ell)}\left(\tanh\frac{x_1}{2},\ldots,\tanh\frac{x_{2\ell}}{2}\right)\tanh{\frac{x_{2\ell +1}}{2}}
\end{align}
Now we observe that the elementary symmetric polynomial of degree $j$ in $\ell$ variables can be decomposed as $\sigma_j^{(\ell)}(a_1,\ldots, a_\ell)= \sigma_j^{(\ell-1)}(a_1,\ldots, a_{\ell-1})+\sigma_{j-1}^{(\ell-1)}(a_1,\ldots, a_{\ell-1})\,a_\ell$, hence
\begin{align}
&\sum_{j=0}^{\ell -1} \left[\sigma_{2j+1}^{(2\ell)}(\tanh{\frac{x_1}{2}},\dots,\tanh{\frac{x_{2\ell}}{2}}) +  \sigma_{2j}^{(2\ell)}\left(\tanh\frac{x_1}{2},\ldots,\tanh\frac{x_{2\ell}}{2}\right)\tanh{\frac{x_{2\ell +1}}{2}}\right] \nonumber \\
=&\sum_{j=0}^{\ell -1}\sigma_{2j+1}^{(2\ell+1)}(\tanh{\frac{x_1}{2}},\dots,\tanh{\frac{x_{2\ell+1}}{2}})\,.
\end{align}
Thus the sum of \eqref{app first term} and \eqref{app second term} yields \eqref{mainfodd}. In an analogous way it is possible to prove \eqref{mainf} starting from \eqref{mainfodd}.

\end{document}